\documentclass[11pt,a4paper]{article}
\pdfoutput=1	

\usepackage{jcappub}
\usepackage{amsmath}
\usepackage{amsfonts}
\usepackage{amssymb}
\usepackage{mathtools}
\usepackage{graphics}
\usepackage{graphicx}
\usepackage{url}
\usepackage{xcolor}

\usepackage[utf8]{inputenc}

\usepackage{hyperref}

\newcommand{\e}[1]{\times10^{#1}}

\title{Measuring the supernova unknowns  at the next-generation neutrino telescopes through the diffuse neutrino background}
\author[a]{Klaes M\o ller,}
\emailAdd{xgb609@alumni.ku.dk}
\author[a]{Anna M. Suliga,}
\emailAdd{zgs313@alumni.ku.dk}
\author[a,b]{Irene Tamborra,}
\emailAdd{tamborra@nbi.ku.dk}
\author[a]{and Peter B.~Denton}
\emailAdd{denton@nbi.ku.dk}

\affiliation[a]{Niels Bohr International Academy, Niels Bohr Institute, University of Copenhagen, Blegdamsvej 17, 2100, Copenhagen, Denmark}
\affiliation[b]{DARK, Niels Bohr Institute, University of Copenhagen, Juliane Maries Vej 30, 2100, Copenhagen, Denmark}

\abstract{The detection of the diffuse supernova neutrino background (DSNB) will  preciously contribute to gauge the properties of the core-collapse supernova population. We estimate the DSNB event rate in the next-generation neutrino detectors, Hyper-Kamiokande enriched with Gadolinium,  JUNO, and DUNE. The determination of the supernova unknowns through the DSNB will be heavily driven by Hyper-Kamiokande, given its higher expected event rate, and complemented by DUNE that will help in reducing the parameters uncertainties. Meanwhile, JUNO will be sensitive to the DSNB signal over the largest energy range. A joint statistical analysis of the expected  rates in 20 years of data taking from the above detectors  suggests that we will be sensitive to the local supernova rate at most at a $20-33\%$ level. A non-zero fraction of supernovae forming black holes will be confirmed at a 90\% CL, if the true value of that fraction is $\gtrsim20\%$. On the other hand, the DSNB events show extremely poor statistical sensitivity to the nuclear equation of state and mass accretion rate of the progenitors forming black holes.} 

\begin{document}

\maketitle

\section{Introduction} \label{sec:introduction}
A supernova emits about $10^{58}$ neutrinos during its core collapse~\cite{Janka:2017vcp,Mirizzi:2015eza}. The flux of neutrinos coming from all core-collapse supernovae (CC-SN) exploding somewhere in our Universe is dubbed the Diffuse Supernova Neutrino Background (DSNB), see Refs.~\cite{Beacom:2010kk,Lunardini:2010ab,Mirizzi:2015eza} for recent reviews. 

The DSNB is an isotropic and stationary flux whose detection will provide us with a glimpse on the global SN population, paving the way to a new era for neutrino astronomy. In principle, through neutrinos, we may be able to independently constrain the local SN rate, estimate the fraction of SNe evolving into black holes (BH-SN, the otherwise called ``failed'' SNe), the overall features of the nuclear equation of state (EoS) governing the SN progenitors, the average neutrino emission properties, and possibly the explosion mechanism~\cite{Lunardini:2009ya,Horiuchi:2017qja,Nakazato:2015rya,Nakazato:2013maa}. 
 The DSNB is also sensitive to the neutrino oscillation physics, especially to the matter effects enhancing the flavor conversion probability~\cite{Lunardini:2012ne,Ando:2004hc,Chakraborty:2008zp,Chakraboty:2010sz}.

Recent one-dimensional hydrodynamical SN simulations seem to suggest that the fraction of BH-SNe can reach up to $40\%$ of the whole SN population~\cite{Ertl:2015rga,Sukhbold:2015wba}. Although such results need to be confirmed by self-consistent studies employing multi-dimensional hydrodynamical simulations, they hint towards a fraction of BH-SNe larger than assumed in the past. Those theoretical findings are in agreement with results obtained by direct searches hunting for the disappearance of red supergiants~\cite{Adams:2016hit,Adams:2016ffj,Gerke:2014ooa,Kochanek:2008mp}. If confirmed, the existence of a large fraction of BH-SN progenitors will contribute to alleviate the tension between the measured and the theoretically estimated SN rate~\cite{Horiuchi:2011zz} and enhance the high energy tail of the DSNB energy spectrum. 

The DSNB has not been detected yet. The most stringent upper limit has been placed by Super-Kamiokande~\cite{Bays:2011si,Zhang:2013tua,Zhang:2015zla} and is about a factor of two above the current theoretical predictions.  The enrichment of Super-Kamiokande with Gadolinium (Gd)~\cite{Beacom:2003nk,Horiuchi:2008jz,Watanabe:2008ru,Xu:2016cfv}
is going to drastically enhance the signal-over-background ratio and it is  likely to  finally lead to the DSNB detection with about a 3 $\sigma$ significance after 10 years of data taking, see e.g.~Ref.~\cite{Sekiya:2016xji}. 

After the awaited DSNB detection with the Super-Kamiokande-Gd project, future generation neutrino detectors will allow us to explore the DSNB astrophysical unkowns.
In fact detectors, such as the water Cherenkov detector Hyper-Kamiokande possibly enriched with Gd~\cite{Hyper-Kamiokande:2016dsw} and the Jiangmen Underground Neutrino Observatory (JUNO)~\cite{An:2015jdp} with liquid scintillator, will be sensitive to the DSNB with large statistics. Slow liquid scintillator detectors have been recently proposed as another promising option to this purpose~\cite{Wei:2016vjd,Priya:2017bmm}.
While the above detectors will be mainly sensitive to antineutrinos, the upcoming Deep Underground Neutrino Experiment (DUNE) with liquid Argon may detect the DSNB signal in the neutrino channel, although with less statistics~\cite{Acciarri:2016crz}. 
  
In this work, for the first time, we explore the statistical chances to  constrain some of the SN unknowns through the DSNB neutrinos by taking into account degeneracies of the DSNB signal among the model parameters. 
The manuscript is organized as follow. In Sec.~\ref{sec:nusignal}, we introduce the neutrino emission properties for CC-SN and BH-SN progenitors as well as the expected distribution of SNe on cosmological distances, and forecast the expected DSNB signal. In Sec.~\ref{sec:detectors}, we outline the features of the next-generation neutrino detectors involved in our analysis (Hyper-Kamiokande with Gd, JUNO, and DUNE) and estimate the expected event rate in each of these detectors. In Sec.~\ref{sec:significancetest}, the probability of exploring the astrophysical parameters affecting the DSNB, such as the fraction of BH-SN progenitors, the average nuclear equation of state (EoS), the BH-SN mass accretion rate, and the local SN rate are outlined. A generalization of the above analysis is then presented in Sec.~\ref{sec:analysis}. In Sec.~\ref{sec:conclusions}, conclusions on our findings and perspectives are reported. Finally, details on the statistical analysis and degeneracies among the model parameters are discussed in Appendix~\ref{sec:appendix1}. 

\section{Diffuse supernova neutrino background} \label{sec:nusignal}
In this Section, we introduce the neutrino emission properties employed to model the CC-SN and BH-SN populations as well as the cosmic SN rate  and its uncertainties. After convolving the signal with the relevant flavor conversion physics, the expected DSNB signal is presented. 

\subsection{Neutrino emission properties for core-collapse and failed supernovae} \label{sec:nusignal1}

To simulate the entire SN population, we employ the neutrino emission properties from one-dimensional spherically symmetric hydrodynamical simulations of the SN core collapse from the Garching group~\cite{Garc:SN,Mirizzi:2015eza}. In order to investigate the variability of the expected signal as a function of the SN progenitor mass and the nuclear equation of state (EoS) for the standard CC-SN population, we use a set of progenitors with mass $9.6$ and $27\ M_\odot$, each progenitor with Lattimer and Swesty EoS with nuclear incompressibility modulus $K=220$~MeV (LS220 EoS)~\cite{Lattimer:1991nc} and the SFHo hadronic EoS (SFHo EoS)~\cite{Steiner:2012rk} for a total of four models. The BH-SN progenitors are instead modeled through two $40 M_\odot$ progenitors: models s40c and s40s7b2c, respectively with ``slow'' and ``fast'' black hole formation~\cite{Mirizzi:2015eza}; both runs were computed with LS220 EoS, but have different mass accretion rates. In Sec.~\ref{sec:Snrate}, details on the parameterization of the SN population mass range through the above six SN progenitors will be described. 

\begin{figure}[t]
\centering
\includegraphics[width=1.\textwidth]{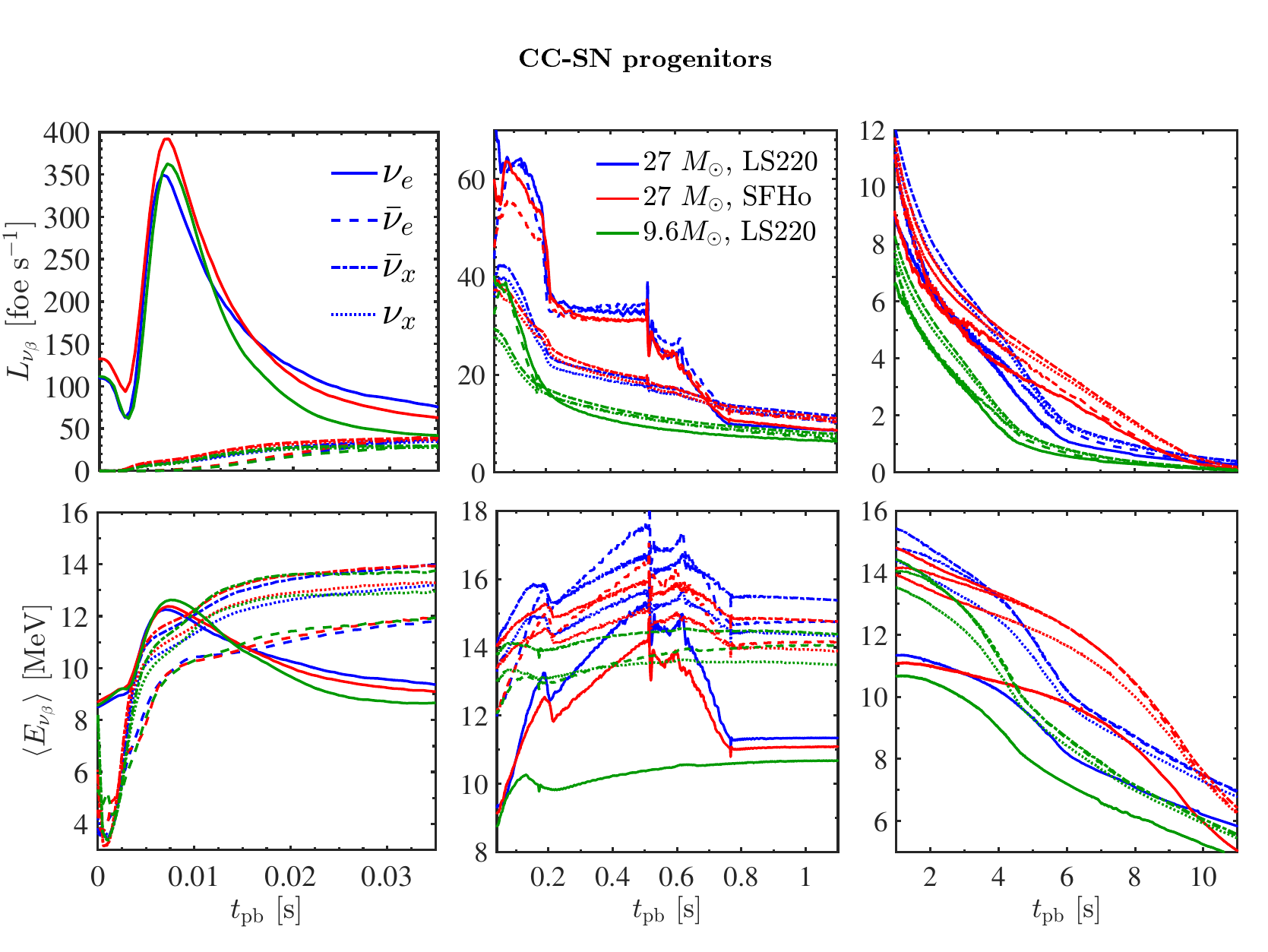}
 \caption{Neutrino luminosities (on the top) and mean energies (on the bottom) in the observer frame as a function of the post-bounce time from 1D spherically symmetric hydrodynamical SN simulations of the 
 progenitors with mass $9.6$ (LS220 EoS, in green) and $27\ M_\odot$ (with LS220 EoS in blue and SFHo EoS in red, respectively)~\cite{Garc:SN,Mirizzi:2015eza}. The $\nu_e$ properties are drawn with continuous lines, the $\bar{\nu}_e$ ones with dashed lines, the $\nu_x=\nu_\mu, \nu_\tau$ with dotted lines, and the $\bar{\nu}_x$ with dash-dotted lines.
 }
 \label{lum_meane_CCSN}
\end{figure}

To provide an idea of the variability range of the CC-SN neutrino properties due to the progenitor mass and nuclear EoS dependence, the top panel of Fig.~\ref{lum_meane_CCSN} shows the neutrino luminosity $L_{\nu_\beta}$ as a function of the post-bounce time for each flavor $\nu_\beta=\nu_e, \bar{\nu}_e, \nu_x$ and $\bar{\nu}_x$ (with $x=\mu,\tau$) for the $27 M_\odot$ progenitor, with LS220 and SFHo EoS, and the $9.6 M_\odot$ CC-SN with LS220 EoS. The bottom panel of Fig.~\ref{lum_meane_CCSN} shows the correspondent average neutrino energies $\langle E_{\nu_\beta}\rangle$.
The neutrino signal in Fig.~\ref{lum_meane_CCSN} is divided in three time windows corresponding to the neutronization burst, the accretion phase, and the Kelvin-Helmholtz cooling phase, from left to right respectively. We refer the interested reader to Sec.~2.4.6 of Ref.~\cite{Mirizzi:2015eza} for a discussion on the EoS dependence of the neutrino emission properties. 

\begin{figure}[t]
\centering
\includegraphics[width=1.\textwidth]{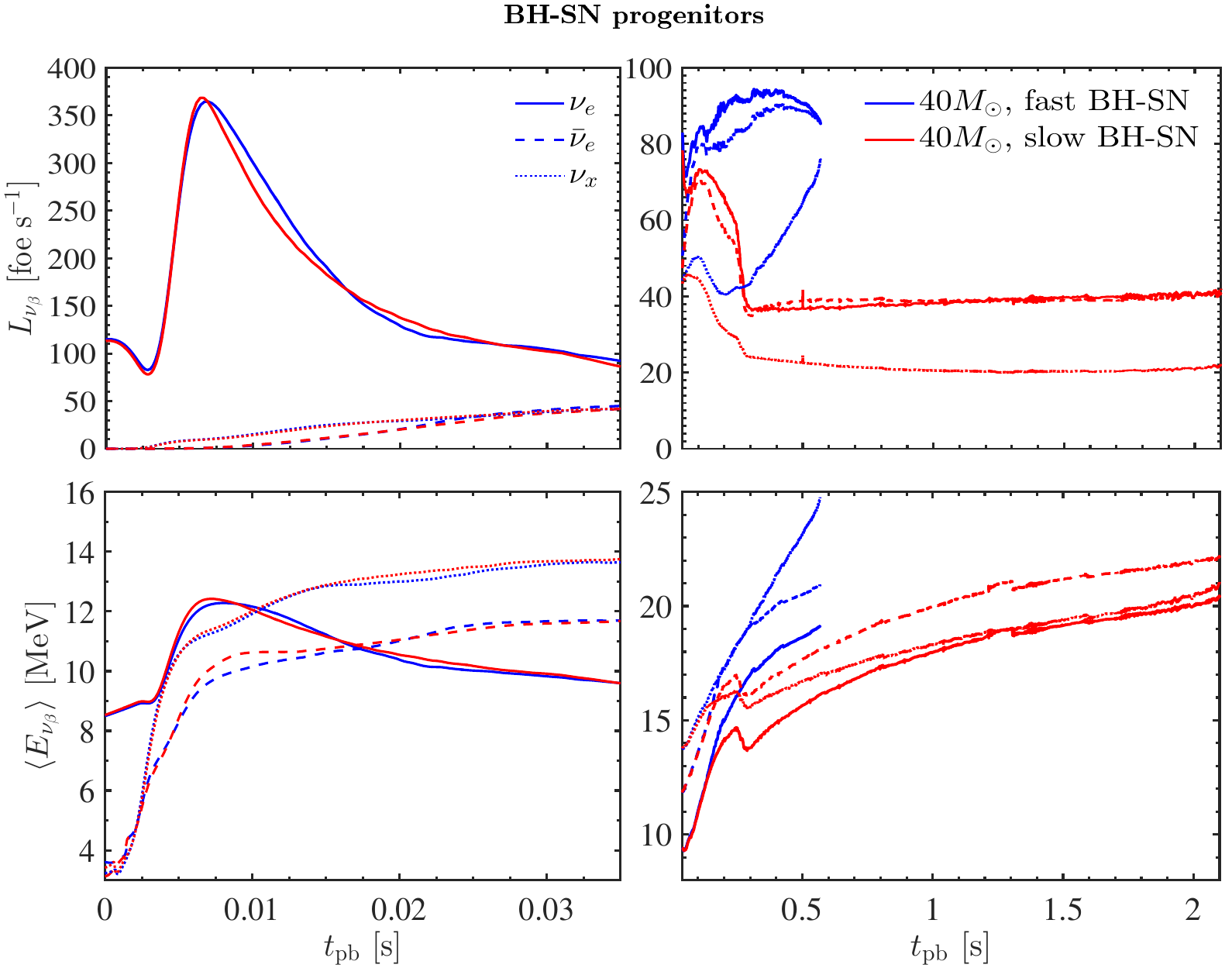}
 \caption{Neutrino luminosities (on the top) and mean energies (on the bottom) in the observer frame as a function of the post-bounce time for  1D spherically symmetric hydrodynamical simulations of a BH-SN
 progenitor with mass $M= 40\ M_\odot$ for the fast (s40s7b2c, in blue) and for the slow (s40c, in red) black hole formation cases~\cite{Garc:SN,Mirizzi:2015eza}. The $\nu_e$ properties are drawn with continuous lines, the $\bar{\nu}_e$ ones with dashed lines, the $\nu_x=\nu_\mu, \nu_\tau$ with dash-dotted lines, and the $\bar{\nu}_x$ with dotted lines. 
}
 \label{lum_meane_BHSN}
\end{figure}
For comparison, Fig.~\ref{lum_meane_BHSN} shows the neutrino properties as a function of time for the BH-SN cases with the ``fast'' and ``slow'' black hole formation rate. 
The duration of the neutrino signal is quite different between the s40c and s40s7b2c progenitors, and shorter than the CC-SN cases shown in Fig.~\ref{lum_meane_CCSN} because of the sudden black hole formation. While it takes only 0.57~s after bounce for model s40s7b2c to form a black hole (fast BH-SN), the gravitational instability occurs after 2.1 s for the s40c model (slow BH-SN). This is because the collapsing stellar core has a mass accretion rate that is lower after the composition shell interface has passed the stalled shock  in the s40c model (see Sec.~2.5 of Ref.~\cite{Mirizzi:2015eza} for details). The time-dependent differences among the neutrino properties are most prominent for the $\nu_x$ flavors.

At each time $t_{\mathrm{pb}}$ after the core bounce, the neutrino differential flux at a distance $d$ is parametrized as
\begin{equation}
F^0_{\nu_\beta}(E,t_{\mathrm{pb}})=\frac{L_{\nu_\beta}(t_{\mathrm{pb}})}{4 \pi d^2} \frac{\phi_{\nu_\beta}(E,t_{\mathrm{pb}})}{\langle E_{\nu_\beta}(t_{\mathrm{pb}})\rangle}\ ,
\end{equation}
where $\phi_{\nu_\beta}(E,t_{\mathrm{pb}})$ is the neutrino energy distribution~\cite{Keil:2002in,Tamborra:2012ac}
\begin{equation}
\label{eq:phi}
\phi_{\nu_\beta}(E,t_{\mathrm{pb}})=\xi_{\nu_\beta}(t_{\mathrm{pb}}) \left(\frac{E}{\langle E_{\nu_\beta}(t_{\mathrm{pb}})\rangle}\right)^{\alpha_\beta(t_{\mathrm{pb}})} \exp\left[-\frac{(1+\alpha_\beta(t_{\mathrm{pb}})) E}{\langle E_{\nu_\beta}(t_{\mathrm{pb}})\rangle}\right]\ ;
\end{equation}
the pinching factor $\alpha_\beta(t_{\mathrm{pb}})$ satisfies the following relation
\begin{equation}
\frac{\langle E_{\nu_\beta}(t_{\mathrm{pb}})^2\rangle}{\langle E_{\nu_\beta}(t_{\mathrm{pb}})\rangle^2}=\frac{2+\alpha_\beta(t_{\mathrm{pb}})}{1+\alpha_\beta(t_{\mathrm{pb}})}\ ,
\end{equation}
while $\xi_{\nu_\beta}(t_{\mathrm{pb}})$ is a normalization factor ($\int dE\, \phi_{\nu_\beta}(E,t_{\mathrm{pb}}) = 1$).

\subsection{Neutrino flavor conversions} \label{sec:oscillations}
Close to the neutrino-sphere, in the proximity of the neutrino decoupling region, $\nu$--$\nu$ interactions are not negligible given the high-neutrino density~\cite{Chakraborty:2016yeg,Mirizzi:2015eza,Duan:2010bg}. While during the neutronization burst, neutrino self-interactions effects are suppressed, fast pairwise conversions as well as ordinary slow $\nu$--$\nu$ interactions may affect the later stages of the neutrino signal~\cite{Chakraborty:2016yeg,Mirizzi:2015eza,EstebanPretel:2008ni,Tamborra:2017ubu,Izaguirre:2016gsx}. Reference~\cite{Lunardini:2012ne} presented an estimation of the DSNB which included a full multi-angle study of the neutrino self-interactions. Since $\nu$--$\nu$ interactions induce energy-dependent signatures in the energy spectrum varying with the post-bounce time, any effect due to the dense neutrino medium is averaged out and has a negligible impact on the DSNB~\cite{Lunardini:2012ne}. In this paper, we will also neglect any modification of the neutrino fluxes due to fast pairwise conversions~\cite{Tamborra:2017ubu,Izaguirre:2016gsx,Sawyer:2015dsa,Sawyer:2005jk}. 

While neutrinos propagate through the SN envelope, they interact with the electron background giving rise to the well known Mikheev-Smirnov-Wolfenstein (MSW) enhanced flavor conversions~\cite{Mikheev:1986if,1978PhRvD..17.2369W,1985YaFiz..42.1441M}. For each of the adopted SN progenitor models introduced in Sec.~\ref{sec:nusignal1}, we computed the neutrino and antineutrino survival probabilities numerically by solving the neutrino kinetic equations of motion and employing time dependent matter profiles from Ref.~\cite{Garc:SN}. We found that, even for the BH-SN progenitors, flavor conversions in matter are fully adiabatic and are well reproduced by the analytical approximation~\cite{Dighe:1999bi,Chakraboty:2010sz} for both normal (NO) and inverted (IO) mass orderings: 
 \begin{equation}
F_{\nu_e,\mathrm{NO}} = s^2_{12} P_H (F_{\nu_e}^0-F_{\nu_x}^0) + F_{\nu_x}^0\ \mathrm{and}\ F_{\bar{\nu}_e,\mathrm{NO}} = c^2_{12} (F_{\bar{\nu}_e}^0-F_{\nu_x}^0) + F_{\nu_x}^0\ , 
\end{equation}
\begin{equation}
F_{\nu_e,\mathrm{IO}} = s^2_{12} (F_{\nu_e}^0-F_{\nu_x}^0) + F_{\nu_x}^0\ \mathrm{and}\ F_{\bar{\nu}_e,\mathrm{IO}} = c^2_{12} P_H (F_{\bar{\nu}_e}^0-F_{\nu_x}^0) + F_{\nu_x}^0\ , 
\end{equation}
where $s_{12}=\sin \theta_{12}$ and $c_{12}=\cos \theta_{12}$. The probability $P_H$ is the effective jump probability due to the atmospheric mass splitting. Given the value of $\theta_{13}$, $P_H \simeq 0$~\cite{Dighe:1999bi}; while $s^2_{12} \simeq 0.297$~\cite{Capozzi:2017ipn}. 
Note that if fast conversions should induce flavor equilibration for a certain time window of the neutrino signal then the resulting DSNB will be in between the most extreme cases considered here of MSW effects in NO and IO. 

In this work we will assume that the neutrino mass ordering will have been determined from terrestrial experiments at the time of the DSNB detection with the next generation detectors, see e.g.~Ref.~\cite{Patterson:2015xja} for a recent review on the topic. 
Hence, we will only focus on the determination of the astrophysical unknowns affecting the DSNB and assume that the mass ordering is fixed. 

\subsection{Cosmic core-collapse supernova rate} \label{sec:Snrate}
The SN rate describes the redshift ($z$) and progenitor distribution of the entire core-collapse population. 
The SN rate is proportional to the star-formation rate ($\dot{\rho}_\star$) through the following relation
\begin{equation}
R_{\mathrm{SN}}(z,M)=\frac{\eta(M)}{\int_{0.5 M_\odot}^{125 M_\odot} dM M \eta(M)} \dot{\rho}_\star(z)\ ,
\end{equation} 
where $\eta(M)$ is the initial mass function that we assume to follow the Salpeter law ($\eta(M) \propto M^{-2.35}$)~\cite{1955ApJ...121..161S}. Variations of the DSNB signal due to changes of the Salpeter scaling law within its error bars have been found to induce minimum variations on the predicted DSNB~\cite{Horiuchi:2017qja}. 

A piecewise parameterization of the star-formation rate is assumed~\cite{Yuksel:2008cu,Strolger:2015kra,Madau:2014bja}: 
\begin{equation}
\dot{\rho}_\star(z) \propto \left[(1+z)^{p_1 k} + \left(\frac{1+z}{5000}\right)^{p_2 k}+\left(\frac{1+z}{9}\right)^{p_3 k}\right]^{1/k}\ ,
\end{equation}
with $k=-10$, $p_1=3.4$, $p_2=-0.3$, and $p_3=-3.5$. 
The local SN rate is normalized such that $\int_{8 M_\odot}^{125 M_\odot} dM R_{\mathrm{SN}}(0,M) = 1.25 \pm 0.5 \times 10^{-4}$~Mpc$^{-3}$~yr$^{-1}$~\cite{Lien:2010yb}. Note that the value adopted for the uncertainty in $R_{\mathrm{SN}}(0)$ is such that includes most of the estimated local rates from astronomical surveys~\cite{Strolger:2015kra,Mattila:2012zr,Taylor:2014rlo,Botticella:2011nd,Li:2010kc,Li:2010kd,Lien:2010yb,Cappellaro:2015qia}. In fact the variability range for the local SN rate considered here intrinsically takes into account contributions from BH-SNe and CC-SNe.

The SN rate includes the fraction of CC- and BH-SNe. To that purpose, we define the fraction of BH-SNe as
\begin{equation}
f_{\mathrm{BH-SN}} = \frac{\int_{\Lambda_\mathrm{BH-SN}} \eta(M) dM}{\int_{8 M_\odot}^{125 M_\odot} \eta(M) dM}\ ,
\end{equation}
where $\Lambda_\mathrm{BH-SN}$ represents the domain in the progenitor mass range where one expects to have BH-SN progenitors; from here $f_{\mathrm{CC-SN}} = 1 - f_{\mathrm{BH-SN}}$. 

In order to study the DSNB sensitivity to $f_{\mathrm{BH-SN}}$, we consider three different scenarios:
\begin{itemize}
\item[-] $f_{\mathrm{BH-SN}} = 9\%$. This corresponds to the assumption that all stars with $M \ge 40 M_\odot$ evolve into BH-SNe~\cite{Woosley:2002zz,Pejcha:2014wda}. This is a conservative scenario where the fraction of SNe leading to black hole formation is small. The progenitor mass distribution has been modeled by assuming the $9.6 M_\odot$ ($27 M_\odot$) neutrino signal as benchmark signal for the progenitors in the range $[8 M_\odot, 15 M_\odot]$ ($[15 M_\odot, 40 M_\odot]$), and the $40 M_\odot$ signal as reference for the BH-SN progenitors with $M > 40 M_\odot$. 
\item[-] $f_{\mathrm{BH-SN}} = 21\%$. In this case, following Refs.~\cite{Ertl:2015rga,Raithel:2017nlc}, progenitors with $M > 22-25 M_\odot$ were assumed to evolve into BH-SNe. The progenitor distribution has been modeled by assuming the $9.6 M_\odot$ ($27 M_\odot$) neutrino signal as benchmark signal for the progenitors in the range $[8 M_\odot, 15 M_\odot]$ ($[15 M_\odot, 22 M_\odot] + [25 M_\odot, 27 M_\odot]$), and the $40 M_\odot$ signal as reference for the BH-SN progenitors in the range $[22 M_\odot, 25 M_\odot] + [27 M_\odot, 125 M_\odot]$.
 \item[-] $f_{\mathrm{BH-SN}} = 41\%$. This is based on the findings of Ref.~\cite{Sukhbold:2015wba,Hidaka:2016zei} and it is still well within the observational constraints~\cite{Adams:2016hit}. It represents our most optimistic case for the DSNB detection as it will be clear from the next Section. The progenitor mass distribution has been modeled by assuming the $9.6 M_\odot$ signal as benchmark for the progenitors in the range $[8 M_\odot, 15 M_\odot]$, and the $40 M_\odot$ signal as reference for the BH-SN progenitors with $M > 15 M_\odot$.
 \end{itemize}
 Concerning the last scenario,  BH-SN progenitors with $M > 15 M_\odot$ may have a neutrino signal slightly different than the one of the $40 M_\odot$ progenitor adopted here as well as they might greatly differ on their accretion rate. However, given the lack of BH-SN models with sophisticated neutrino transport, we here rely on a conservative assumption.
Note that $f_{\mathrm{BH-SN}}$ could be a function of $z$. Given the data currently available, such a dependence is very uncertain, see e.g.~Ref.~\cite{Mattila:2012zr}. However, we verified that a redshift dependence of $f_{\mathrm{BH-SN}}$ along the lines of the one proposed in Ref.~\cite{Mattila:2012zr} would have a negligible impact on the DSNB with respect to the one computed by adopting a constant $f_{\mathrm{BH-SN}}$. 

\subsection{Expected diffuse supernova neutrino background} \label{sec:dsnbtheory}
The DSNB is defined as
\begin{gather}
\begin{aligned}
\Phi_{\nu_\beta}(E)={}&\frac{c}{H_0} \int_{8 M_\odot}^{125 M_\odot} dM \int_0^{z_\mathrm{max}} dz \frac{R_{\mathrm{SN}}(z,M)}{\sqrt{\Omega_M (1+z)^3 + \Omega_\Lambda}}\\
&\times\left[f_{\mathrm{CC-SN}} F_{\nu_\beta,\mathrm{CC-SN}}(E^\prime,M) + f_{\mathrm{BH-SN}} F_{\nu_\beta,\mathrm{BH-SN}}(E^\prime,M)\right]\ ,
\end{aligned}
\end{gather}
where $c$ is the speed of light, $\Omega_M=0.3$ and $\Omega_\Lambda=0.7$ are the matter and dark energy cosmic energy densities, and $H_0 = 70$ km s$^{-1}$ Mpc$^{-1}$~\cite{Rao:2005ab}, and $z_{\mathrm{max}}=5$. The neutrino energy $E^\prime$ is the energy at the production site at redshift $z$ related to the energy at the Earth by $E^\prime = E (1+z)$. $F_{\nu_\beta}(E^\prime,M)$ is the time-integrated energy spectrum for each progenitor mass $M$.

\begin{figure}[t]
\centering
\includegraphics[width=1.\textwidth]{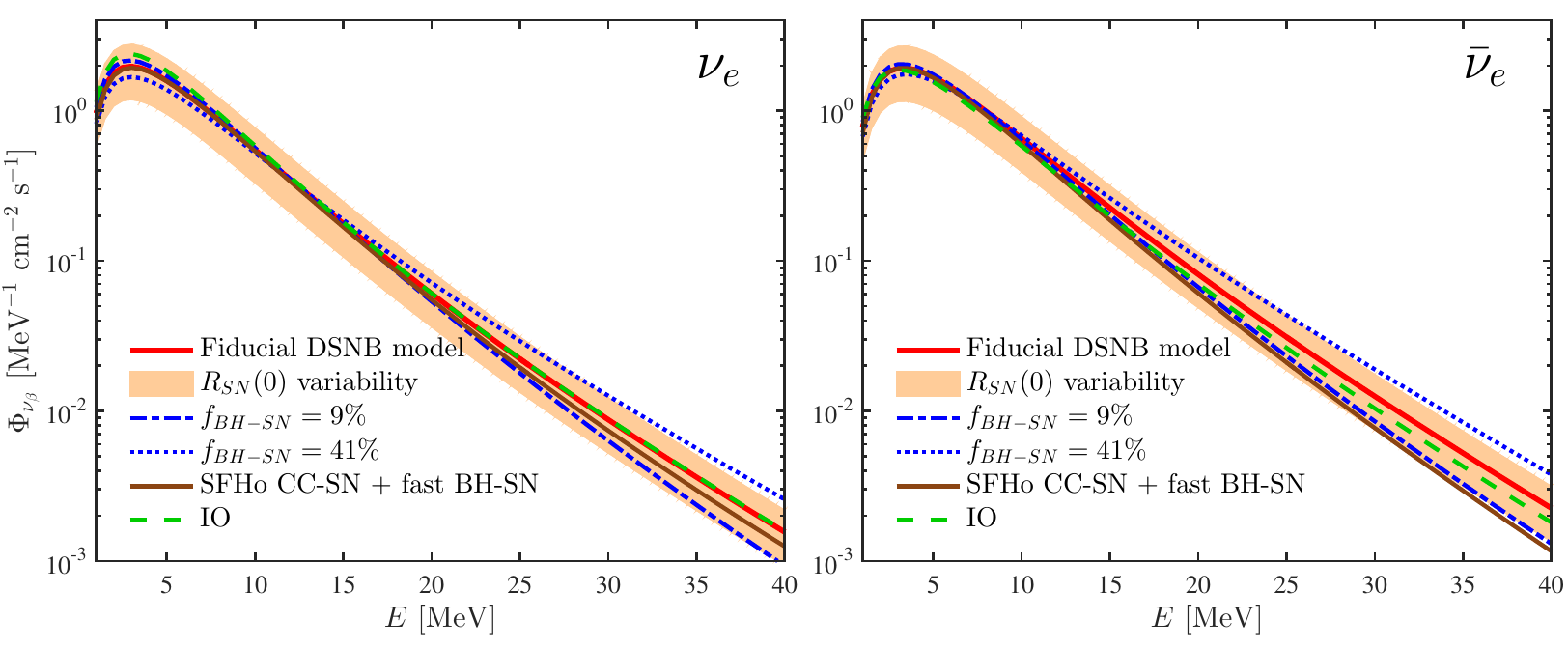}
 \caption{DSNB as a function of the neutrino energy for $\nu_e$ (on the left) and $\bar{\nu}_e$ (on the right). The ``fiducial DSNB model'' is plotted in red; it has been obtained for the slow BH-SN case and LS220 EoS for the CC-SN, $R_{\mathrm{SN}}(0)=1.25 \times 10^{-4}$~Mpc$^{-3}$~yr$^{-1}$, $f_{\mathrm{BH-SN}} = 21\%$, and NO. The orange band takes into account the $R_{\mathrm{SN}}(0)$ uncertainty ($R_{\mathrm{SN}}(0) \in [1.25-0.5, 1.25+0.5] \times 10^{-4}$~Mpc$^{-3}$~yr$^{-1}$). Variations of the fiducial DSNB model with respect to 
$f_{\mathrm{BH-SN}}$ are shown in blue, while the maximum variation of the fiducial model due to the CC-SN EoS and BH-SN accretion rate  is plotted in brown. The green dashed line represents the DSNB for IO. 
}
 \label{DSNB_theo}
\end{figure}

Figure~\ref{DSNB_theo} shows the predicted DSNB as a function of the neutrino energy for $\nu_e$ on the left and $\bar{\nu}_e$ on the right. In the following, unless otherwise stated, we will rely on our so-called ``fiducial DSNB model.'' The latter is defined as the DSNB computed for $f_{\mathrm{BH-SN}} = 21\%$, slow BH-SN and LS220 EoS for the CC-SN progenitors, $R_{\mathrm{SN}}(0)=1.25 \times 10^{-4}$~Mpc$^{-3}$~yr$^{-1}$, and NO. In order to gauge the DSNB variability range, the orange band represents the variation of the fiducial DSNB model due to the local SN rate uncertainty, i.e.~$R_{\mathrm{SN}}(0) \in [1.25-0.5, 1.25+0.5] \times 10^{-4}$~Mpc$^{-3}$~yr$^{-1}$. 
The variability range of the DSNB due to
$f_{\mathrm{BH-SN}}$ with respect to the fiducial model is shown in blue for $f_{\mathrm{BH-SN}}=9, 41 \%$. The maximum variation of the fiducial DSNB model with respect to the CC-SN EoS and BH-SN accretion rate is plotted in brown (for the SFHo EoS for the CC-SN and the fast BH-SN case)\footnote{Other combinations of the CC-SN EoSs and the fast vs.~slow BH-SN cases are possible than the ones considered here. However, we tested that those are the ones giving the largest DSNB variation.} and it is larger in the antineutrino channel. 

From Eq.~\ref{eq:phi} and Figs.~\ref{lum_meane_CCSN} and \ref{lum_meane_BHSN}, one can see that the BH-SNe have hotter energy spectra (i.e., higher $\langle E_{\nu_\beta}\rangle$) than the CC-SNe  especially at late post-bounce times. This leads to a higher energy tail for $F_{\nu_\beta,\mathrm{BH-SN}}$. 
In the DSNB, when integrating over all progenitor masses, this feature is most prominent in the case of the slow BH-SNe due to their longer emission period. As a consequence, a larger $f_{\mathrm{BH-SN}}$ results in a flatter spectrum and a higher flux at high energies, as also shown in e.g.~Refs.~\cite{Lunardini:2009ya, Lien:2010yb}.
The fact that the brown line (SFHo CC-SN + fast BH-SN) in Fig.~\ref{DSNB_theo} gives a lower flux at high energies is primarily due to the short emission period of the fast BH-SN.
The green dashed line in Fig.~\ref{DSNB_theo} represents the DSNB in IO, and $\bar{\nu}_e$'s are mostly sensitive as expected. Note, however, that this trend may change as $f_{\mathrm{BH-SN}}$ increases as the relative ratio between the resultant electron and non-electron fluxes in the DSNB  changes (see discussion in Sec.~\ref{sec:significancetest}). 

\section{Forecast of the DSNB event rate at future generation neutrino detectors} \label{sec:detectors}
In this Section, we present the DSNB event rate and the backgrounds in the next-generation neutrino detectors included in our analysis: Hyper-Kamiokande (Gd), JUNO, and DUNE. We will also discuss the optimal energy range for the DSNB detection for each facility. Given current uncertainties on when the detectors will be ready to start taking data, how much active time each of them will have, as well as what will be the final volumes, we choose to work in the best scenario option, where DSNB signal rates in Hyper-Kamiokande  (Gd), JUNO, and DUNE were calculated for 20 yrs of data taking each. Differences in their start times and respective efficiencies will have only a small effect on our conclusions.

\subsection{Hyper-Kamiokande enriched with Gadolinium}
Hyper-Kamiokande~\cite{Hyper-Kamiokande:2016dsw} is an upcoming water Cherenkov neutrino detector to be built in Japan as a successor of the Super-Kamiokande detector. The currently preferred configuration consists of two cylindrical tanks, each of them with 187 kton of fiducial volume. The main detection channel is the inverse beta decay (IBD) process
\begin{equation}
\bar{\nu}_e + p \rightarrow n + e^+\ .
\label{eq:ibd}
\end{equation}

It is currently under discussion the possibility that the nominal configuration of the detector may be modified by enriching it with Gadolinium sulfate, according to the outcome of the Super-Kamiokande Gadolinium Project currently under development~\cite{Sekiya:2016xji,Beacom:2003nk}. It is estimated that the addition of $0.1\%$ of GdCl$_3$ to water will reduce the invisible muon background by a factor of about 5~\cite{Hyper-Kamiokande:2016dsw}. Given the improvements due to the Gd enrichment on the DSNB detection~\cite{Beacom:2003nk}, we will assume that this is the case in this manuscript unless otherwise stated. The detector efficiency is therefore $\epsilon = 67\%$ (being $90\%$ the neutron capture efficiency and $74\%$ the event selection efficiency). 
 
The event rate is expected to be
\begin{equation}
N(E_{e^+}) = \epsilon N_t \ \int dE \Phi_{\bar{\nu}_e}(E)\sigma_{\mathrm{IBD}}(E, E_{e^+})\ ,
\end{equation}
with $N_t = 2.5 \times 10^{34}$ the number of targets and $\sigma_{\mathrm{IBD}}$ the IBD cross-section~\cite{Strumia:2003zx}. 
\begin{figure}[h]
\centering
\includegraphics[width=0.8\textwidth]{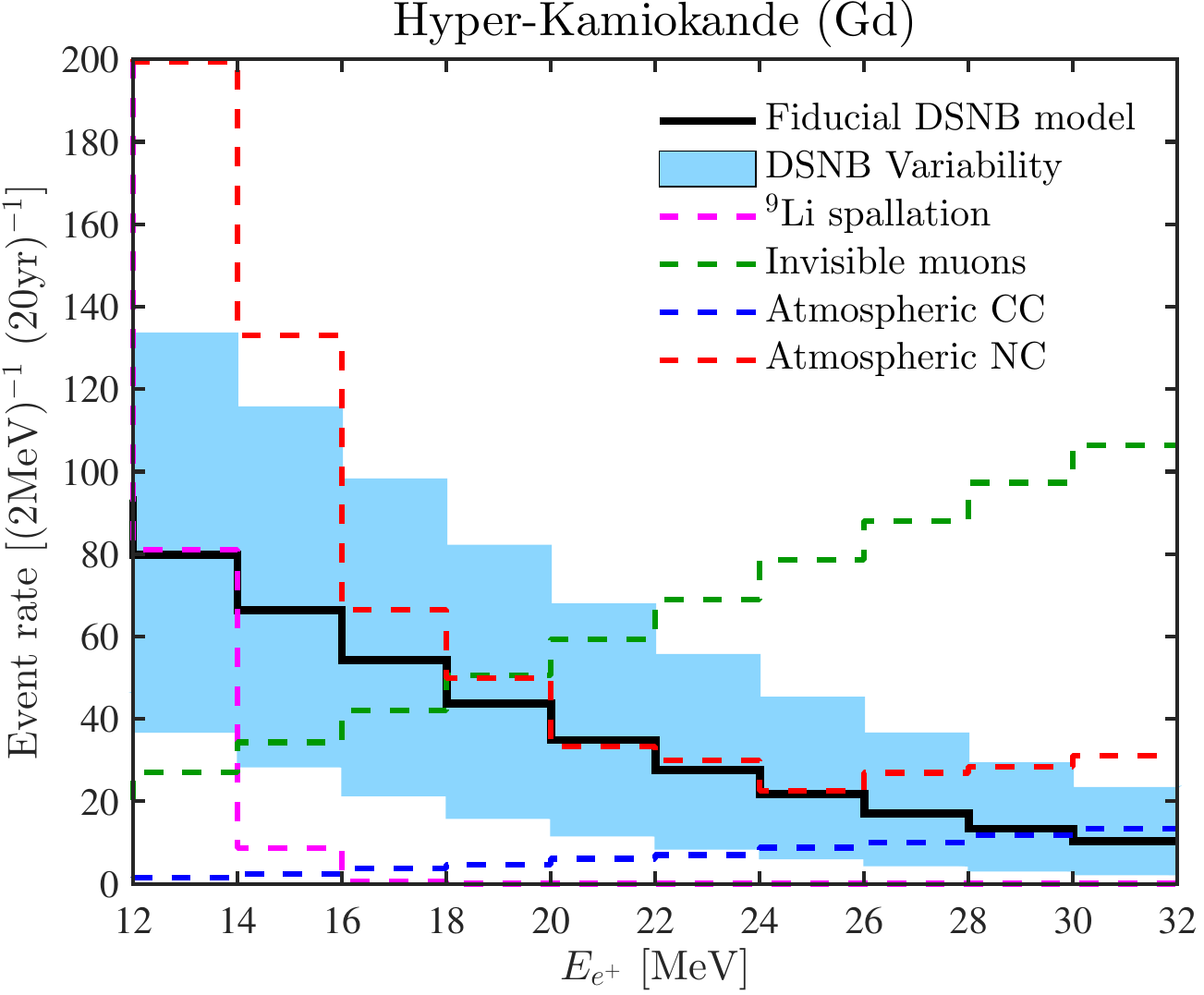}
 \caption{Expected DSNB event rate as a function of the positron energy in Hyper-Kamiokande (Gd) for 20 years of data taking. The event rate for the DSNB fiducial model is plotted in black, the maximum expected model variability within the theoretical uncertainties (see Sec.~\ref{sec:nusignal}) is marked by the blue band. The background due to atmospheric CC (NC) events is plotted in blue (red), the background due to invisible muons is plotted in green and the one due to $^9$Li spallation is shown in magenta. The optimal detection window is between 12 and 24 MeV, assuming full efficiency in removing the NC atmospheric background.}
 \label{DSNB_HK}
\end{figure}
\begin{figure}[h]
\centering
\includegraphics[width=0.8\textwidth]{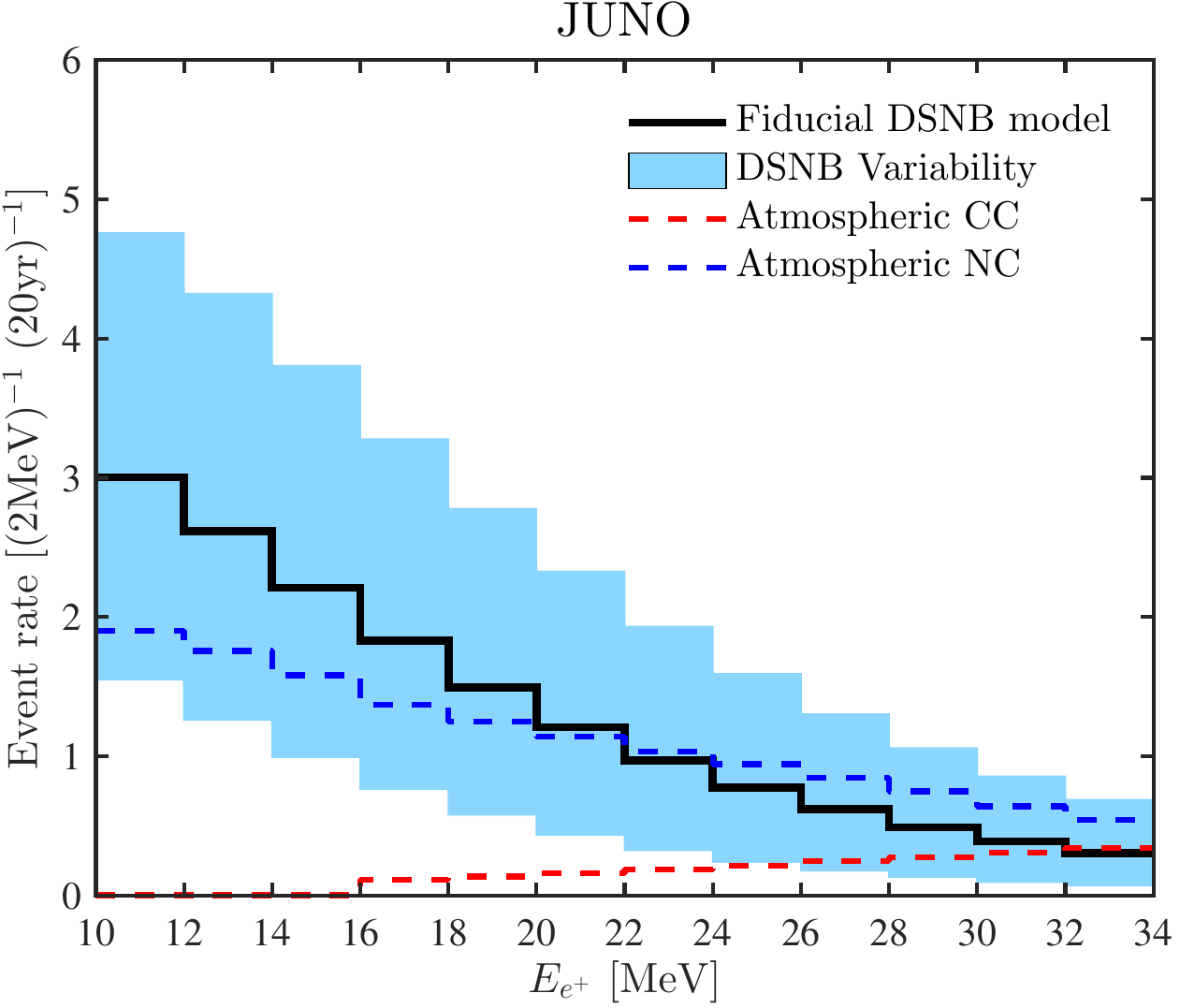}
\caption{Expected DSNB even rate as a function of the positron energy in JUNO for 20 years of data taking. The event rate for the DSNB fiducial model is plotted in black, the maximum expected model variability within the theoretical uncertainties (see Sec.~\ref{sec:nusignal}) is marked by the blue band. The background due to atmospheric CC (NC) events is plotted in red (blue). The DSNB signal is going to be clearly distinguishable from the background in the positron energy range between 10 and 24 MeV.}
\label{DSNB_JUNO}
\end{figure}

The residual backgrounds affecting the DSNB are the atmospheric $\nu_e$ and $\bar{\nu}_e$ charged current (CC) events coming from the interactions of cosmic rays with the atmosphere, the atmospheric $\nu_\mu$ and $\bar{\nu}_\mu$ giving rise to the so-called Michel spectrum, and $^9$Li which is the only spallation product with accompanying neutrons. The neutral current (NC) atmospheric events, given by de-excitation $\gamma$-rays induced by NC quasi elastic scattering~\cite{Kunxian:2015ymr} should be an unavoidable background; however, there might be ways to reach a full tagging discrimination~\cite{Beacom:privcomm}. Hence, we will neglect this background in the combined statistical analysis developed in the next Section. The backgrounds in Hyper-Kamiokande (Gd) are implemented according to Refs.~\cite{Hyper-Kamiokande:2016dsw,Kunxian:2015ymr}. 

Figure~\ref{DSNB_HK} shows the expected DSNB event rate in Hyper-Kamiokande (Gd) in 20 years of data taking. The DSNB variability due to the theoretical uncertainties discussed in the previous Section is marked in blue. The various backgrounds discussed above are represented by the dashed curves. As it is evident from Fig.~\ref{DSNB_HK}, assuming that the atmospheric NC background can be neglected, the optimal positron energy range for the DSNB detection is 12--24 MeV.  In fact, above 12 MeV one avoids the reactor neutrino background and reduces the influence from $^9$Li spallation. The upper bound for the optimal detection window has been chosen so to optimize the significance of the $\chi^2$ test (see Sec.~\ref{sec:significancetest}); the inclusion of bins with higher energies reduces the outcome of the $\chi^2$ test for most DSNB models.
The theoretical uncertainties are larger than the statistical uncertainties, which will allow for parameter estimation as shown in Section \ref{sec:significancetest}.

Of course, the existing Super-Kamiokande currently enriched with Gd will be also sensitive to the DSNB measurement with their 22.5 ktons fiducial volume~\cite{Bays:2011si}, although with less statistics. Since the detector technologies are very similar,  we focus our analysis on Hyper-Kamiokande; we note that including Super-Kamiokande to our forecast requires only a straightforward rescaling of our Hyper-Kamiokande rates.

\subsection{JUNO}
The Jiangmen Underground Neutrino Observatory (JUNO) is a 20~kton liquid scintillator detector to be built in Jiangmen, China~\cite{An:2015jdp}. The detector should consist of a central tank filled with linear alkylbenzene. 
The central detector is immersed in a water Cherenkov detector surrounded by a muon tracker in order to reduce the backgrounds, resulting in a fiducial volume of 17 kton ($N_t=1.2 \times 10^{33}$). 
The event rate is
\begin{equation}
N(E_{e^+}) = \epsilon N_t \int dE \Phi_{\bar\nu_e}(E)\sigma_{\mathrm{IBD}}(E, E_{e^+})\ ,
\end{equation}
with $\epsilon$ varying for the DSNB signal and the backgrounds as from Table 5-1 of Ref.~\cite{An:2015jdp} (see  next paragraph). 

The backgrounds affecting the DSNB detection are discussed at length in Sec.~5.3 of Ref.~\cite{An:2015jdp}. The ones that cannot be avoided are the reactor $\bar{\nu}_e$ flux that is larger than the DSNB spectrum for $E_{e^+}< 9$~MeV. We therefore set 10 MeV as the minimum observed positron energy. The atmospheric CC $\bar{\nu}_e$ spectrum cannot be completely tagged, similarly to the NC atmospheric flux. The fast neutrons produced by cosmic muons decaying outside the detector also constitute a background; however, since most of these events should happen near the edge of the detector, this background can be strongly reduced by considering a smaller fiducial volume (17 kton). A significant reduction of the backgrounds can be achieved, if pulse shape discrimination techniques are applied~\cite{An:2015jdp}. In this work, we take into account the pulse shape discrimination with an efficiency varying as a function of the background type, as reported in Table 5-1 of Ref.~\cite{An:2015jdp}. 

Figure~\ref{DSNB_JUNO} shows the expected DSNB event rate to be observed in JUNO for 20 year of data taking. The atmospheric CC and NC backgrounds are plotted in red and blue respectively. 
The optimal positron energy window for the DSNB detection in JUNO is  10--24~MeV.  The lower energy limit is set by the reactor background and the upper limit (24~MeV) has been chosen to maximize the signal over background ratio. Although, JUNO will have less statistics than Hyper-Kamiokande (Gd), see Figs.~\ref{DSNB_HK} and \ref{DSNB_JUNO}, it will be sensitive to the DSNB signal in a wider energy range and thus be a valuable tool to examine the DSNB spectral shape  and increase the statistics in combined analysis.
The statistical uncertainties for JUNO will be of a similar size to the theoretical ones.
So, while JUNO will not be able to make strong statements on the DSNB properties itself, it will contribute when combined with the other experiments.

\subsection{DUNE}
The Deep Underground Neutrino Experiment (DUNE) contains a 40 kton Liquid Argon (LAr) detector to be built in South Dakota~\cite{Acciarri:2016crz,Acciarri:2016ooe,Acciarri:2015uup}. The current planning foresees the construction of four individual LAr time projection chambers, each with $\sim10$~kton LAr. In this work, we assume four identical time projection chambers resulting in a 40~kton fiducial volume. 
The main neutrino detection channel is
\begin{equation}
\nu_e + ^{40}\mathrm{Ar} \rightarrow e^- + ^{40}\mathrm{K}^\ast\ .
\label{eq:larint}
\end{equation}

The event rate is expected to be
\begin{equation}
N(E) = \epsilon N_t \Phi_{\nu_e}(E)\sigma_{\mathrm{Ar}}(E)\ ,
\end{equation}
with $N_t = 6.02\times 10^{32}$ the number of target Argon atoms, and $\sigma_{\mathrm{Ar}}$ the $\nu$--Ar cross-section~\cite{Martinez}.
DUNE is expected to have a trigger efficiency of about $90\%$ and a reconstruction efficiency of $96\%$~\cite{Ankowski:2016lab,DUNEp}; hence we assume a detection efficiency $\epsilon = 86\%$.

The backgrounds affecting the DSNB detection in DUNE are still under investigation. We here assume that the backgrounds will be comparable to the ones affecting the DSNB detection in the ICARUS detector~\cite{Cocco:2004ac}. 
We remind the interested reader to Sec.~4 of Ref.~\cite{Cocco:2004ac} for a detailed discussion on the backgrounds. The ones that cannot be eliminated or tagged are the hep and $^8$B solar neutrinos. Since the end tail of the hep flux is at $19$~MeV, we consider the DSNB event rate above that energy in our analysis. 
The atmospheric $\nu_e$ flux is another irreducible background affecting the high energy tail of the neutrino spectrum. As we will see in the next Section, the atmospheric background sets the maximum neutrino energy up to which the DSNB events are easily distinguishable from the background. 

\begin{figure}[t]
\centering
\includegraphics[width=0.8\textwidth]{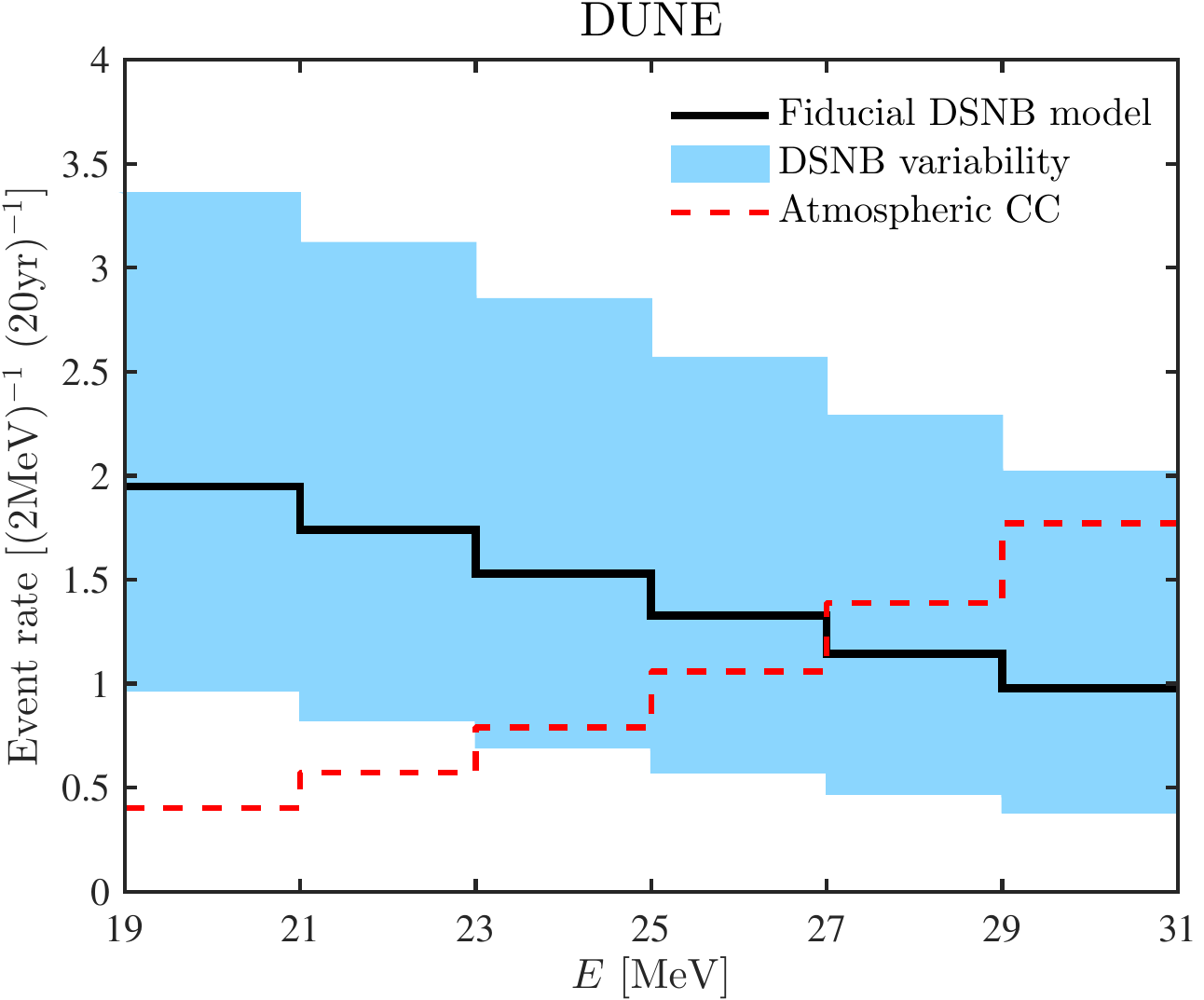}
\caption{Expected DSNB event rate as a function of the neutrino energy in DUNE for 20 years of data taking. The event rate for the DSNB fiducial model is plotted in black, the maximum expected model variability within the theoretical uncertainties (see Sec.~\ref{sec:nusignal}) is marked by the blue band. The background due to atmospheric CC events is plotted in red. The optimal DSNB detection window is in the range between 19 and 29 MeV.}
\label{DSNB_DUNE}
\end{figure}
Figure~\ref{DSNB_DUNE} shows the expected signal as a function of the neutrino energy.  The optimal DSNB detection window in DUNE is between 19 and 29 MeV. The lower limit in this window is determined by the irreducible solar neutrino background which dominates the event rate at $E<19 \mathrm{MeV}$. The upper limit has been picked to be such to maximize the significance of the $\chi^2$ test. Although the DUNE event rate is smaller than the ones expected in Hyper-Kamiokande (Gd) and JUNO (see Figs.~\ref{DSNB_HK} and \ref{DSNB_JUNO}), DUNE provides complementary information with respect to the other detectors considered in this work in that it is mainly sensitive to neutrinos, while the other two detectors are sensitive to antineutrinos.
Similarly to JUNO, the statistical uncertainties for DUNE will be comparable to the theoretical ones.
Note that DUNE will  not be able to make strong statements on the DSNB properties itself, but it will contribute when combined with the other experiments.

\section{Detectability prospects} \label{sec:significancetest}
In this Section, we discuss the chances of pinning down the the fraction of BH-SN or the local SN rate, or average information on the BH-SN accretion rate  by using the DSNB neutrinos. By minimizing with respect to the other model parameters,  a joint statistical analysis of the neutrino event rate from Hyper-Kamiokande (Gd), JUNO, and DUNE is developed. The role of the various detectors in determining such parameters is also outlined. A generalization of this analysis will be presented in the next Section.
While the energy spectra in the detectors are roughly linear within the energy regions of interest, we perform a complete statistical analysis by calculating the $\chi^2$ test statistics in each bin for each detector.

\subsection{Significance test: Fraction of black hole forming supernovae} \label{sec:chitest}
To estimate the errors expected on the measurement of the astrophysical parameters affecting the DSNB, such as $f_{\mathrm{BH-SN}}$, the $\chi^2$ pull method has been implemented \cite{Fogli:2002pt}. In order to determine the uncertainty on $f_{\mathrm{BH-SN}}$, we need to minimize over  the local SN rate ($R_{\mathrm{SN}}(0)$), the atmospheric flux uncertainty ($x$), the uncertainty on the cross-section ($y$), and additionally the nuclear EoS--BH-SN mass accretion rate ($z$). 

The statistical analysis has been developed by using a 2 MeV binning of the event rate observed in the various detectors (see Sec.~\ref{sec:detectors}). The number of adopted energy bins is 6 in Hyper-Kamiokande in the range 12--24 MeV, 7 in JUNO in the range between 10 and 24 MeV and 5 in DUNE in the range 19--29 MeV, we assume 20 years of data taking for each detector. 

The resultant $\chi^2$ is defined as 
\begin{equation}
\chi^2 = \min_{A}\left(\chi^2_{R_\mathrm{SN}(0)} + \chi^2_{\mathrm{BG}} + \chi^2_{\sigma} + \chi^2_{\mathrm{HK}} + \chi^2_{\mathrm{JUNO}} + \chi^2_{\mathrm{DUNE}}\right)\ ,
\label{eq:chisq}
\end{equation}
where $\chi^2_{R_\mathrm{SN}(0)} = \left[(R_{\mathrm{SN}}(0) - \bar{R}_{\mathrm{SN}}(0))/\Delta_{R_\mathrm{SN}(0)}\right]^2$, $\Delta_{R_\mathrm{SN}(0)}=0.25\e{-4}$ Mpc$^{-3}$ yr$^{-1}$ which is the uncertainty on the local SN rate, and $\bar{R}_{\mathrm{SN}}(0)$ our fiducial value\footnote{Note that all quantities marked with a bar correspond to the fiducial values chosen according to the fiducial DSNB model (see Sec.~\ref{sec:nusignal}).}.
The atmospheric flux is the main source of uncertainty in Hyper-Kamiokande (Gd), JUNO, and DUNE. Hence, 
$\chi^2_{\mathrm{BG}} = \left[x/\Delta_{\mathrm{BG}}\right]^2$
characterizing the background fluctuations: $N_{\mathrm{BG},i} = (1+x) \bar{N}_{\mathrm{BG},i}$ and $\Delta_{\mathrm{BG}} = 20\%$~\cite{Kunxian:2015ymr}. While this conservative estimate of the uncertainty only takes into account the normalization uncertainty of background events, it should cover the range of slope uncertainties over the regions of interests.
For what concerns the uncertainties on the cross-sections, $\chi^2_{\sigma}=\chi^2_{\sigma_\mathrm{Ar}} = \left[y/\Delta \sigma_\mathrm{LAr}\right]^2$ where we have assumed that the cross-section uncertainties are negligible except for DUNE: $\Delta \sigma_\mathrm{LAr} = 15\%$~\cite{Acciarri:2014isz,LArp}. We have taken the uncertainty on the LAr cross-section, $y$, to be independent of energy for simplicity.
 The set of parameters to be marginalized over is $A=\{f_{\mathrm{BH-SN}},R_\mathrm{SN}(0),x,y,z\}$ minus the parameter(s) of interest.

For what concerns the last term in Eq.~\ref{eq:chisq}, the $\chi^2$ is given in terms of a likelihood ratio with Poisson distributions which is necessary since the event rates are $\lesssim1$ for JUNO and DUNE.
Then the $\chi^2$ terms for the experiments are given by,
\begin{equation}
\chi^2_{\mathrm{HK}, \mathrm{JUNO}, \mathrm{DUNE}} = -2 \sum_i\ln\frac{L_{0, E_i}}{L_{1, E_i}}\ ,
\end{equation}
where the sum is over energy bins.
The likelihoods are 
\begin{equation}
L_{0, E_i} \simeq P_i[\lambda=\bar{N}_{\mathrm{BG},i} + N_i(\bar{f}_{\mathrm{BH-SN}}, \bar{R}_{\mathrm{SN}}(0)), k = N_i + N_{\mathrm{BG},i}]\ ,
\end{equation}
and
\begin{equation}
L_{1, E_i} \simeq P_i[\lambda=N_{\mathrm{BG},i} + N_i(f_{\mathrm{BH-SN}}, R_{\mathrm{SN}}(0)), k = N_i + N_{\mathrm{BG},i}]\ .
\end{equation}
$P_i$ is the probability in the energy bin centered on $E_i$ of seeing $k$ events in a Poisson distribution with mean value $\lambda$.
That is, $P_i = \lambda^k \exp({-\lambda})/k!$.
In the case of DUNE, $N_{\mathrm{BG},i} = (1+x) (1+y) \bar{N}_{\mathrm{BG},i}$ and $N_{i} = (1+y) N_i(f_{\mathrm{BH-SN}},R_{\mathrm{SN}}(0))$.

\begin{figure}[t]
\centering
\includegraphics[width=1.\textwidth]{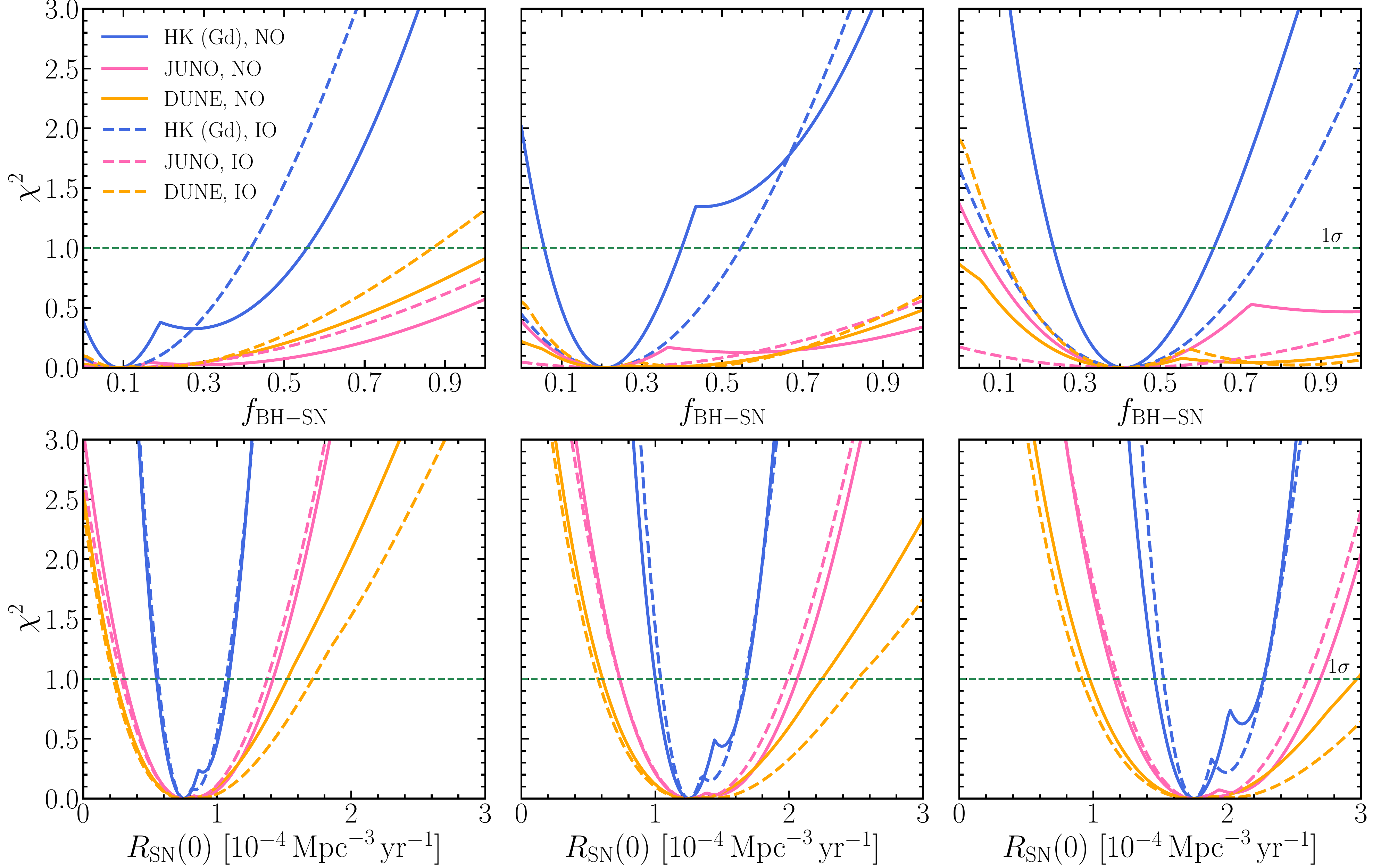}
 \caption{Top panels: Expected $\chi^2$ as a function of the fraction of BH-SNe for Hyper-Kamiokande (Gd), JUNO, and DUNE. The panels from left to right have been obtained for $\bar{f}_{\mathrm{BH-SN}} = 0.09, 0.21$ and $0.41$ in NO (solid lines) and IO (dashed lines). Bottom panels: Expected $\chi^2$ as a function of the local SN rate  (from left to right: $\bar{R}_{\mathrm{SN}}(0) = 0.75, 1.25, 1.75$). Due to its larger statistics, the $\chi^2$ is more constraining for Hyper-Kamiokande (Gd) than for JUNO and DUNE.} 
 \label{fig:fBH_chisq}
\end{figure}

The top panels of Fig.~\ref{fig:fBH_chisq} show the $\chi^2$ as a function of $f_{\mathrm{BH-SN}}$ for $\bar{f}_{\mathrm{BH-SN}} = 0.09$, $0.21$, and $0.41$ from left to right respectively 
 in NO (solid lines) and IO (dashed lines). In each panel, the different curves represent the $\chi^2$ for Hyper-Kamiokande (Gd), JUNO, and DUNE. One can see that due to its larger statistics, the $\chi^2$ is more constraining for Hyper-Kamiokande (Gd) than for JUNO and DUNE. In all cases, the $\chi^2$ is minimized at the correct value of $\bar{f}_{\mathrm{BH-SN}}$ as expected. However, for $\bar{f}_{\mathrm{BH-SN}} = 0.09$ and $0.21$, the $\chi^2$ has another local minimum for Hyper-Kamiokande (Gd) in NO (and similarly for DUNE for $\bar{f}_{\mathrm{BH-SN}} = 0.41$  and JUNO for $\bar{f}_{\mathrm{BH-SN}} = 0.21, 0.41$) because of the degeneracy of the DSNB event rate with respect to the BH-SN accretion rate and $f_{\mathrm{BH-SN}}$. However, the local minimum is not present in IO. We refer the reader to Appendix~\ref{sec:appendix1} for details on the  degeneracy of the model parameters. 

\begin{figure}[t]
\centering
\includegraphics[width=0.95\textwidth]{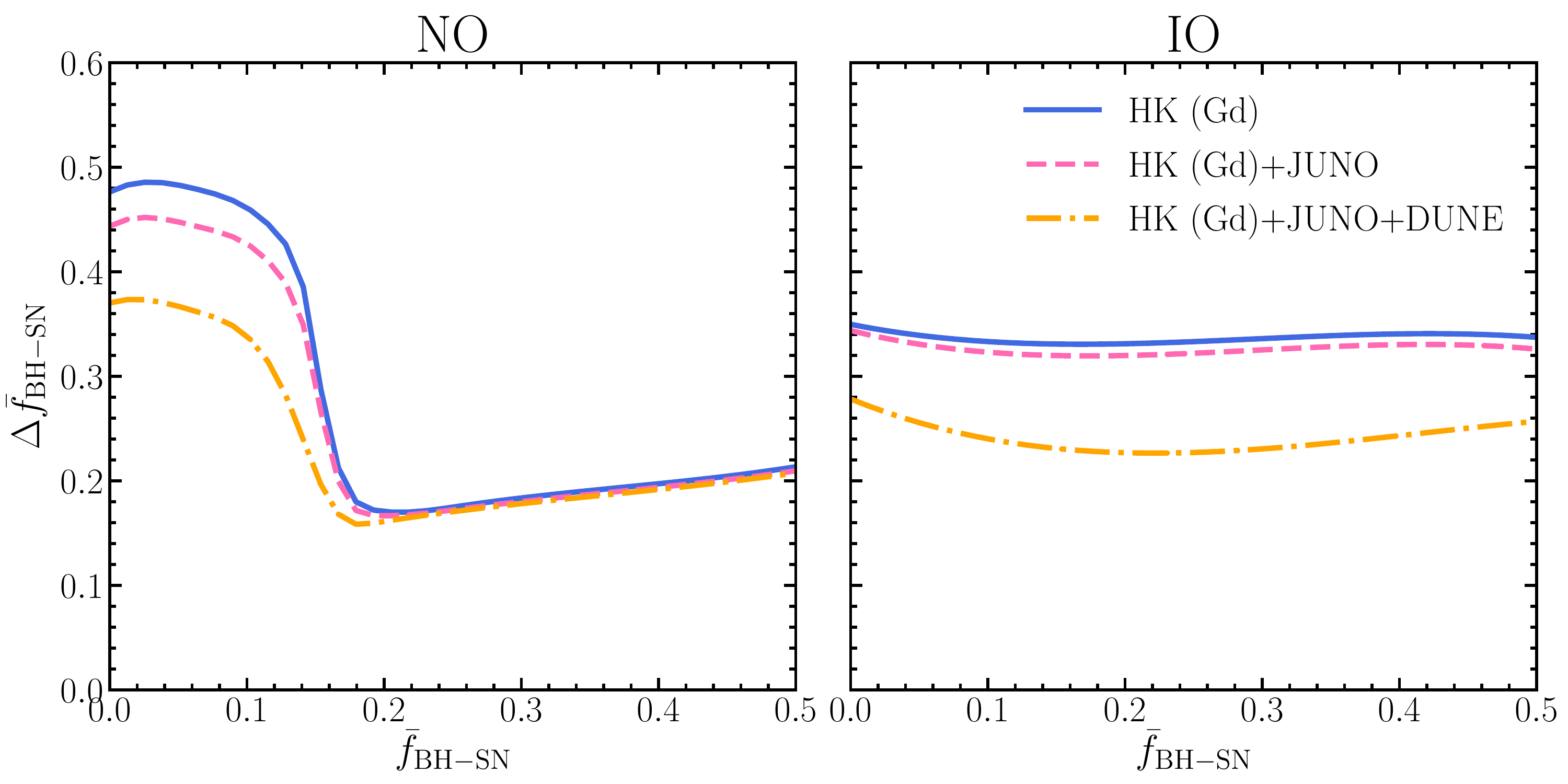}
\includegraphics[width=0.95\textwidth]{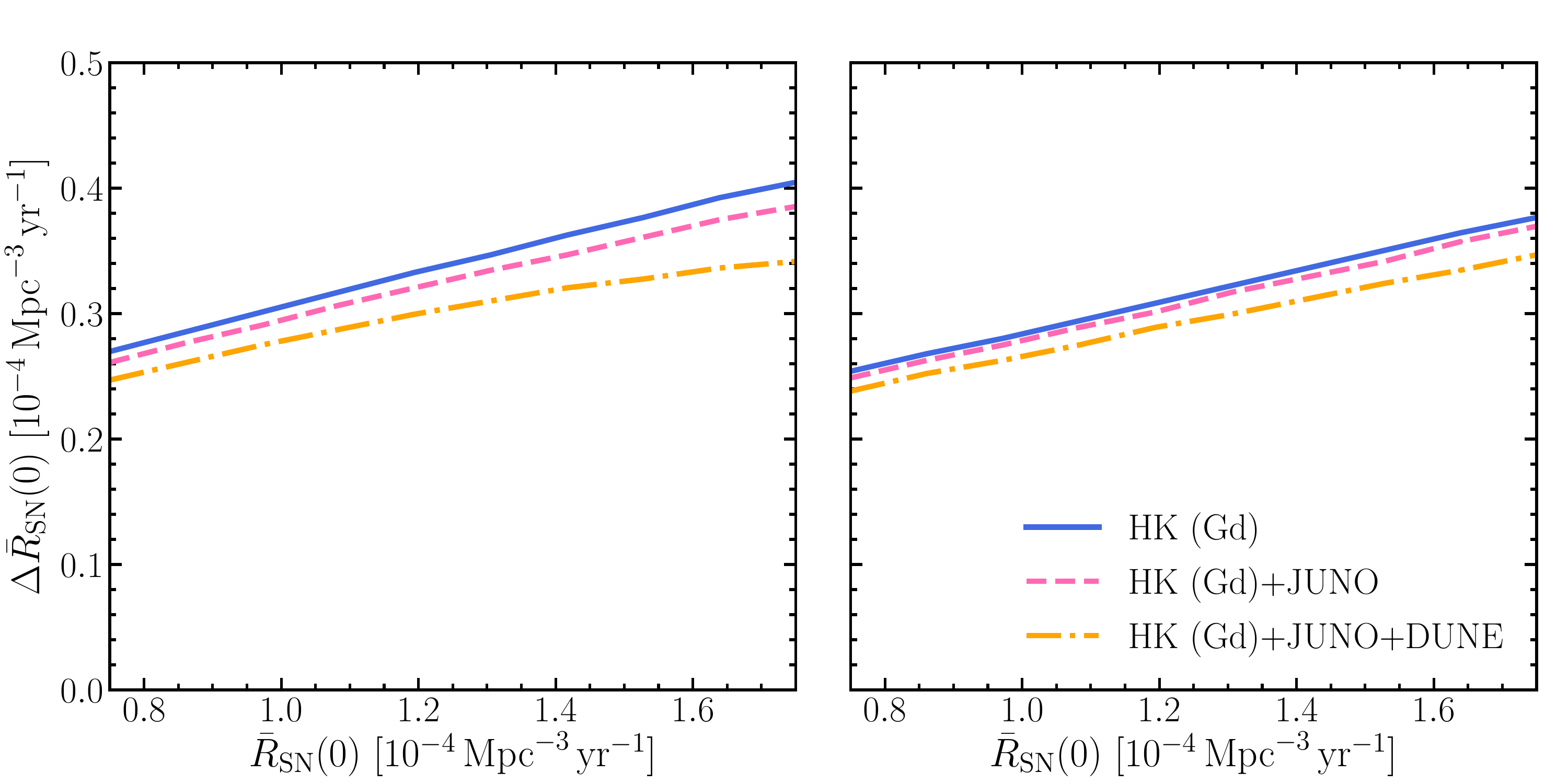}
 \caption{Expected $1 \sigma$ uncertainty for $\Delta \bar{f}_{BH-SN}$ as a function of $\bar{f}_{BH-SN}$ on the top and $\Delta \bar{R}_{\mathrm{SN}}$ as a function of $\bar{R}_{\mathrm{SN}}$ on the bottom in NO (on the left) and IO (on the right). The blue, magenta and yellow curves show the errors obtained by Hyper-Kamiokande (Gd), by combining the event rate observed in Hyper-Kamiokande (Gd) and JUNO, and by   combining the event rate of Hyper-Kamiokande (Gd), JUNO and DUNE, respectively. The other model parameters are kept fixed to our fiducial model, except for $\bar{f}_{BH-SN}=0.14 \pm 0.2$~\cite{Adams:2016hit,Adams:2016ffj} in the bottom panel. Due to the two local minima in the $\chi^2$,  for $\bar{f}_{\mathrm{BH-SN}} \lesssim 0.2$ in the NO, the relative error in its determination is large. A relative error ranging between $33$ and $20\%$ is expected as $\bar{R}_\mathrm{SN}(0)$ increases.} 
 \label{fig:fBH_delta}
\end{figure}
The top panels of Fig.~\ref{fig:fBH_delta} show the error in the determination of $\bar{f}_{\mathrm{BH-SN}}$ as a function of $\bar{f}_{\mathrm{BH-SN}}$ in NO on the left and IO on the right. Due to the presence of two local minima in the $\chi^2$, one can see that for $\bar{f}_{\mathrm{BH-SN}} \lesssim 0.2$ in the NO, the relative error in its determination is large. This happens because of the degeneracies of the expected DSNB flux with the other model parameters (see  Appendix~\ref{sec:appendix1}). The error in the determination of $\bar{f}_{\mathrm{BH-SN}}$  shrinks from $\sim0.35$ down to $\sim0.2$ for the combined analysis (Hyper-Kamiokande (Gd) $+$ JUNO $+$ DUNE) for larger values of $\bar{f}_{\mathrm{BH-SN}}$. In the IO, $\Delta\bar{f}_{\mathrm{BH-SN}} \simeq 0.25$ largely independently of $\bar{f}_{\mathrm{BH-SN}}$. 
Noticeably, the inclusion of the DUNE event rate in the combined statistical analysis can reduce the error in the measurement of $\bar{f}_{\mathrm{BH-SN}}$ up to $20\%$ in NO.  DUNE allows a better precision in the determination of  $\bar{f}_{\mathrm{BH-SN}}$ in IO because the DSNB variations due to $\bar{f}_{\mathrm{BH-SN}}$ are smaller in NO than in IO (see Fig.~\ref{fig:DSNBdeg} which, for the $\nu_e$ case, would be symmetric with respect to NO and IO). This is also proved by the shape of the $\chi^2$ in Fig.~\ref{fig:chisq_parabolas} which is more skewed once DUNE is included for the IO case.

Remarkably, we find that for $\bar f_{\mathrm{BH-SN}} = 0.24$ in NO and $0.35$ in IO, we are able to exclude  $f_{\mathrm{BH-SN}} =0$ at $90\%$ CL by relying on the joint analysis of the observed DSNB event rates only.  Such information will be especially promising if the DSNB constraints will be combined with other measurements of  $\bar f_{\mathrm{BH-SN}}$ coming from electromagnetic surveys.

\subsection{Significance test: Local supernova rate} \label{sec:chitest1}
In order to constrain the local supernova rate, $R_{\mathrm{SN}}(0)$, one needs to proceed as above, but minimizing with respect to the $f_{\mathrm{BH-SN}}$ parameter (i.e., substituting $f_{\mathrm{BH-SN}} \leftrightarrow R_{\mathrm{SN}}(0)$ in Eq.~\ref{eq:chisq}). 
In this case, we fix the model parameters to our fiducial DSNB model except for $\bar{f}_{BH-SN}$; in fact we assume the experimentally measured $\bar{f}_{BH-SN}=0.14 \pm 0.2$ as from Refs.~\cite{Adams:2016hit,Adams:2016ffj}. 

The bottom panels of Fig.~\ref{fig:fBH_chisq} show the $\chi^2$ as a function of $R_{\mathrm{SN}}(0)$. The $\chi^2$ is mostly skewed and it shows a second local minimum for Hyper-Kamiokande (Gd) but it has a negligible  influence on the uncertainty determination than in the case of $\bar f_\mathrm{BH-SN}$ (see Appendix~\ref{sec:appendix1} for details; no degeneracies with respect to the other model parameters appear).

The bottom panel of Fig.~\ref{fig:fBH_delta} shows the $1 \sigma$ uncertainty in the the determination of $\bar{R}_\mathrm{SN}(0)$ for the NO and IO. The variability range for 
$\bar{R}_\mathrm{SN}(0)$ has been fixed according to the uncertainty reported in Sec.~\ref{sec:Snrate}. A relative error ranging between $33$ and $20\%$ is expected as $\bar{R}_\mathrm{SN}(0)$ increases. Since no degeneracy with the other model parameters affect the significance test in this case, the relative error decreases as $\bar{R}_\mathrm{SN}(0)$ increases.

Given its higher statistics, Hyper-Kamiokande (Gd) dominates the statistical uncertainties on the measurement of the local $\bar{R}_{\mathrm{SN}}(0)$. The bottom panel of Fig.~\ref{fig:fBH_delta} shows the average uncertainty ranges between $25$ and $35\%$ according to the value of $\bar{R}_{\mathrm{SN}}(0)$ and it slightly decreases as $\bar{R}_{\mathrm{SN}}(0)$ increases. The addition of JUNO and DUNE improves the uncertainty $\Delta\bar{R}_{\mathrm{SN}}(0)$ by about $5\%$ with respect to the case where Hyper-Kamiokande (Gd) is considered alone. 
 In particular, DUNE allows to pin down the local $\bar{R}_{\mathrm{SN}}(0)$ with higher precision in NO. This can be understood by looking at the shape of the $\chi^2$ in Fig.~\ref{fig:chisq_noparabolas} which is sensibly more skewed once DUNE is included for the NO case.

\subsection{Significance test: Average nuclear equation of state and mass accretion rate for failed progenitors} \label{sec:chitest2}
Although, we could not explore the full parameter space given the lack of input SN models, in this Section we discuss whether the DSNB will show any statistical sensitivity to the average EoS and BH-SN mass accretion rate. We work  under the assumption that the majority of the SN progenitors is well represented by our input SN progenitors. 

Here  we have focused on the  LS220 and SFHo EoSs only, 
under the assumption that they provide a reasonable estimation of the breadth of the EoS parameter space as representative cases for  soft and stiff EoSs, see e.g.~Ref.~\cite{Burgio:2018mcr} for a recent review on the topic. We  estimate the DSNB sensitivity to the EoS. Since the neutrino properties do not differ much  for the LS220 and SFHo EoS cases as shown in Fig.~\ref{lum_meane_CCSN}, we find that the DSNB does not show any sensitivity to the CC-SN EoS (see also Fig.~\ref{fig:fBHtrend} for $\bar{f}_{\mathrm{BH-SN}}=0$).

Figure~\ref{lum_meane_BHSN} shows that the neutrino signal of the BH-SN progenitors is strongly dependent on the  mass accretion rate and this could in turn affect the  DSNB event rate. 
In order to explore the sensitivity of the DSNB to the BH-SN accretion rate, we  calculated the $\Delta\chi^2$  for both mass orderings and for different $\bar{f}_{\mathrm{BH-SN}}$ by minimizing over the remaining parameters in the same fashion as above.
The largest $\Delta\chi^2$ found is $1.8$ ($5.2$) in NO which corresponds to a $1.3\sigma$ ($2.3\sigma$) preference for $\bar{f}_{\mathrm{BH-SN}} = 0.21$ ($0.41$). 
This  is not enough to distinguish between the ``fast'' and ``slow'' BH-SN models, assuming that they are representative of the majority of the BH-SN population.
For most of the parameter space $\Delta\chi^2<1$ and there is no sensitivity to discriminate.

\section{Pinpointing the astrophysical unknowns through the DSNB} \label{sec:analysis}
In the former Section, we have calculated the 1D $\chi^2$ projections for $R_{\mathrm{SN}}(0)$, $f_{\mathrm{BH-SN}}$, and the average EoS--BH-SN accretion rate. In this Section, we will generalize the significance test, by quantifying the uncertainties and by investigating the chances of learning about $\bar{R}_{\mathrm{SN}}(0)$ and $\bar{f}_{\mathrm{BH-SN}}$ simultaneously.

\begin{figure}[t]
\centering
\includegraphics[width=1.\textwidth]{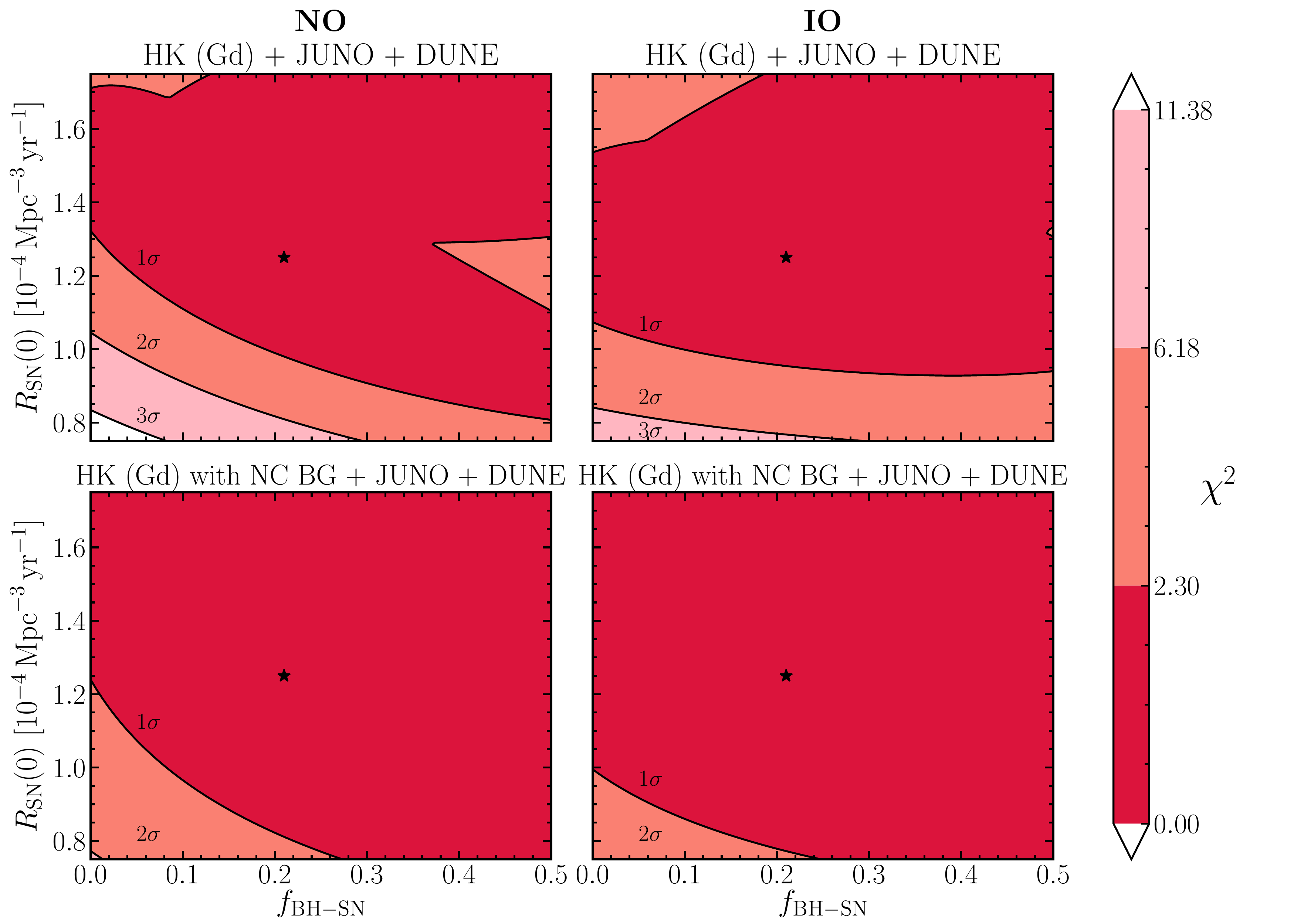}
 \caption{Contour plot of the local $R_{\mathrm{SN}}(0)$ as a function of $f_{\mathrm{BH-SN}}$. The fiducial DSNB model is marked by a star. The upper (bottom) panels refer to the case in which the NC atmospheric background is (not) neglected in Hyper-Kamiokande (Gd). The left (right) panels have been obtained assuming the NO (IO). The three contours indicate the $1, 2$ and $3 \sigma$ confidence levels, respectively for 2 d.o.f. Given the degeneracies among the model parameters, larger values of $R_{\mathrm{SN}}(0)$ or $f_{\mathrm{BH-SN}}$ do not  always correspond to an improved precision in the measurement of the SN unknowns. Moderate precision in the determination of the model parameters can be achieved in IO, if the NC atmospheric background can be neglected in Hyper-Kamiokande (Gd). }  \label{fig:contours}
\end{figure}
The top panels of Fig.~\ref{fig:contours} show a contour plot of the $\chi^2$ for 2 d.o.f.~in the $f_{\mathrm{BH-SN}}$--$R_{\mathrm{SN}}(0)$ plane in the NO on the left and IO on the right. The model parameters corresponding to the fiducial DSNB model are marked with a star in the plot and the three contours indicate the $1$, $2$, and $3 \sigma$ confidence levels, respectively, for 2 d.o.f. The irregular contours, especially visible in the top upper panel, reflect the multiple local minima in the $\chi^2$ (see Appendix~\ref{sec:appendix1} and Sec.~\ref{sec:significancetest}). 

Moderate precision in the determination of the model parameters can be achieved in the IO as it is also clear from Fig.~\ref{fig:fBH_delta} (see Sec.~\ref{sec:chitest}). One can see that, given the degeneracies among the model parameters, larger values of $R_{\mathrm{SN}}(0)$ or $f_{\mathrm{BH-SN}}$ do not directly imply an improved precision in the measurement of these unknowns through the DSNB only (see e.g.~the shape of the contours in the top left panel). Note that for larger values of $\bar{f}_{\mathrm{BH-SN}}$ one should expect a parameter space constrained more tightly (see Fig.~\ref{fig:contours2} in  Appendix~\ref{sec:appendix1}).

For the sake of completeness, the panels on the bottom of Fig.~\ref{fig:contours} have been obtained relaxing the assumption that the NC atmospheric background in Hyper-Kamiokande (Gd) is negligible. As expected, given the lower signal to background ratio, the determination of the physical parameters is more uncertain. 

We worked under the assumption that the neutrino mass ordering will have been determined by terrestrial neutrino experiments at the time of the DSNB detection. However, as shown in this Section, the DSNB is strongly affected by the mass ordering and the DSNB.  At $f_{\mathrm{BH-SN}}=0.21$ the mass ordering cannot be determined since $\Delta \chi^2 = 0.31$, while at $f_{\mathrm{BH-SN}}=0.41$ there will be weak discriminating power at the $1.3 \ \sigma$ ($\Delta \chi^2=1.8$) level.  The DSNB could then still provide an independent hint on this unknown that should be compared with more precise measurements from terrestrial experiments whose current discriminating power between NO and IO is of about $3 \ \sigma$~\cite{Capozzi:2018ubv}.

\section{Conclusions} \label{sec:conclusions}
The diffuse supernova neutrino background (DSNB) represents the cumulative flux of neutrinos emitted from all core-collapse supernovae exploding throughout the Universe.
With the advent of the next-generation large scale neutrino detectors, we will be able to not only detect the DSNB, but measure it and indirectly constrain the astrophysical unknowns of the core-collapse population.

In this work, we forecast the expected DSNB signal for standard core-collapse and black hole forming supernova progenitors, by adopting state-of-the-art inputs for the modeling of the neutrino emission properties. A variability range for the DSNB signal is considered according to the uncertainty on the local supernova rate ($R_{\mathrm{SN}}(0)$), the dependence of the signal on the average nuclear equation of state and the mass accretion rate for the black hole formation, as well as the fraction of the supernova progenitors forming black holes ($f_{\mathrm{BH-SN}}$).

The event rate is then forecasted in the upcoming water Cherenkov detector Hyper-Kamiokande enriched with Gadolinium, the liquid scintillator detector JUNO, and the liquid Argon detector DUNE. Given the different technologies, these detectors will provide complementary information on the fundamental DSNB properties. For the first time, we have developed a joint statistical analysis to investigate the capability of determining the astrophysical parameters affecting the DSNB such as  the local supernova rate,  the fraction of supernovae further evolving into black holes, and the eventual DSNB sensitivity to the mass-accretion rate of the failed progenitors.

Under the assumption that the neutrino mass ordering will be independently established,  the determination of the supernova unknowns through the DSNB will be heavily driven by Hyper-Kamiokande (Gd) given its higher expected event rate. In particular, a fully efficient tagging of the atmospheric  neutral current  background in Hyper-Kamiokande will be instrumental to a better determination of the allowed  $f_{\mathrm{BH-SN}}$--$R_{\mathrm{SN}}(0)$ parameter space. JUNO will be sensitive to the DSNB signal over the largest energy window, allowing a better determination of the spectral features. In combination with Hyper-Kamiokande (Gd), it will allow an improvement in the determination of the model parameters given the higher combined event rate. Although DUNE will have less statistics,  the fact that it is sensitive to the neutrino channel, and therefore complementary to Hyper-Kamiokande (Gd) and JUNO, will help in reducing the parameters uncertainties, especially on $f_{\mathrm{BH-SN}}$.

According to the fraction of the black hole forming progenitors, the DSNB event rate may however show degeneracies among  the model parameters especially if the neutrino mass ordering is normal. However,  for 20 years of data taking, we should be able to pinpoint the local supernova rate with a relative error $\sim20-33\%$.
Remarkably, we will be able to rule out a null fraction of  failed supernovae at 90\% CL by using the DSNB event rates only,  if at least $\mathcal{O}(20\%)$ supernovae form black holes. The uncertainty in the fraction of supernovae that form black holes  can be as large as $\sim 0.4$ in normal mass ordering for  $f_{\mathrm{BH-SN}}\lesssim 0.25$ due to the degeneracies existing among the DSNB model parameters, but it will be $\simeq 0.25$  otherwise. The DSNB shows extremely poor or no sensitivity to the nuclear equation of state and to the mass accretion rate of the failed progenitors, assuming that the majority of them exhibits  a comparable behavior.

Our findings suggest a great potential of the DSNB in pinning down the local supernova rate and the fraction of supernovae forming black holes, despite the degeneracies among the DSNB model parameters. These  encouraging results will be further strengthened once combined with electromagnetic and gravitational wave data from upcoming surveys, shedding light on the global unknowns of the core-collapse supernova population.  

\acknowledgments
We thank John Beacom and Kate Scholberg for insightful discussions, and Eligio Lisi for useful comments on the manuscript. We are also grateful to Thomas Ertl and Hans-Thomas Janka for granting access to the neutrino data of the supernova simulations adopted in this work. PBD and IT acknowledge support from the Villum Foundation (Project No.~13164), and by the Danish National Research Foundation (DNRF91).
PBD thanks the Danish National Research Foundation (Grant No.~1041811001) for support. The work of IT has also been supported by the Knud H\o jgaard Foundation and the Deutsche Forschungsgemeinschaft through Sonderforschungbereich SFB 1258 ``Neutrinos and Dark Matter in Astro- and Particle Physics'' (NDM).

\appendix
\section{Test statistics for the determination of the fraction of failed supernovae and the local supernova rate} \label{sec:appendix1} 
In this Appendix, we discuss the details of the test statistic analysis presented in Secs.~\ref{sec:significancetest} and \ref{sec:analysis}.  In particular, we will focus on the $\chi^2$ analysis for the determination of the fraction of BH-SN and the local SN rate and motivate our results. 
\begin{figure}[b]
\centering
\includegraphics[width=1.\textwidth]{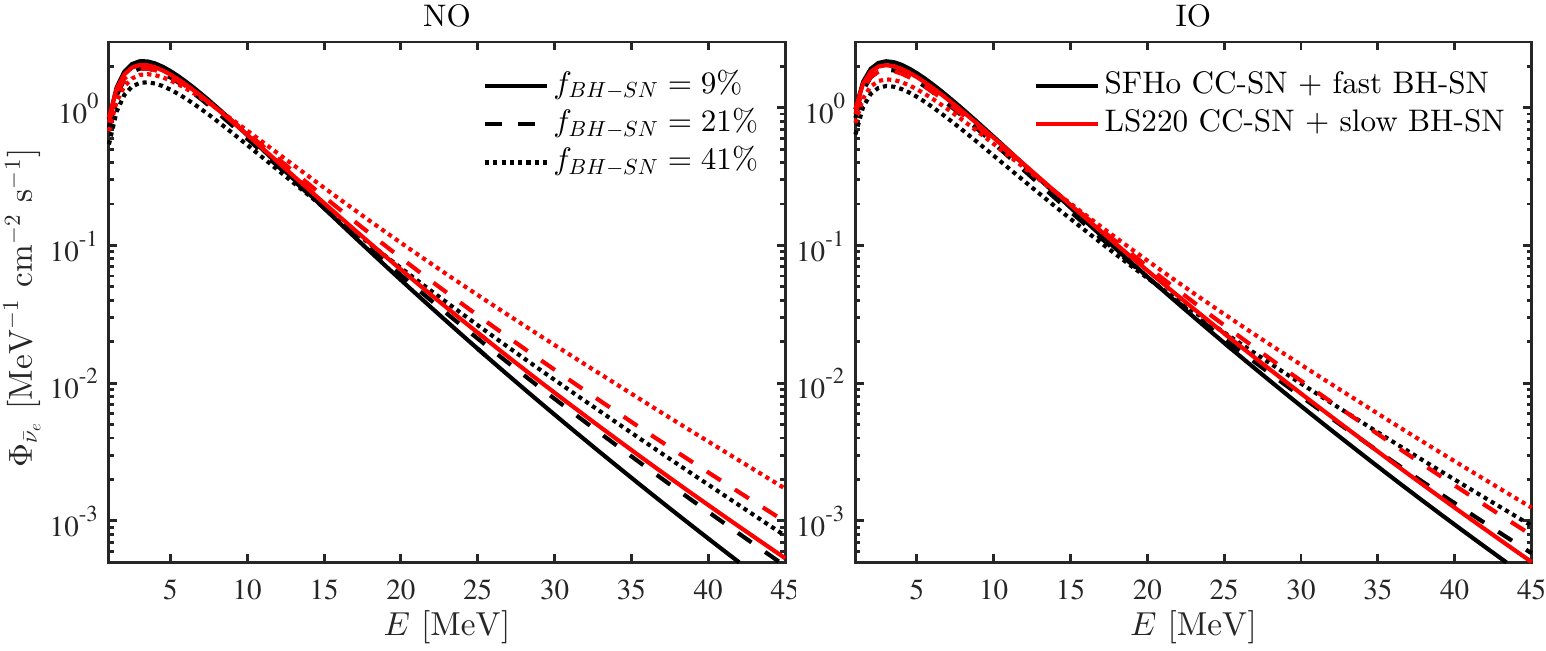}
 \caption{DSNB dependence on the CC-SN EoS--BH-SN mass accretion rate for $f_{\mathrm{BH-SN}} = 9, 21$ and $41\%$ in NO (on the left) and IO (on the right). As $f_{\mathrm{BH-SN}}$ increases, the DSNB fluxes computed by assuming the  SFHo EoS for CC-SN $+$ fast BH-SN cross the ones of the DSNB computed by adopting the LS220 EoS for the CC-SN models $+$ slow BH-SN. The crossing happens roughly at 10 MeV in NO and approximately at 20 MeV in IO. 
 	}
 \label{fig:DSNBdeg}
\end{figure}

The DSNB shows a degeneracy between the CC-SN EoS -- BH-SN mass accretion rate and $f_{\mathrm{BH-SN}}$ as visible in Fig.~\ref{fig:DSNBdeg}; such a degeneracy is dependent on the mass ordering. In fact, as $f_{\mathrm{BH-SN}}$ increases, the DSNB curves obtained by assuming  the SFHo EoS for the CC-SN models $+$ the fast  BH-SN models and the LS220 EoS for CC-SN $+$ the slow BH-SN models cross, creating a degeneracy between these model parameters in the statistical analysis presented in Secs.~\ref{sec:significancetest} and \ref{sec:analysis}. The crossing happens at $\sim$ 10 MeV (15 MeV) in NO and $\sim$ 15 MeV (20 MeV) in IO for the LS220$+$slow (SFHo$+$fast) cases.  

\begin{figure}[t]
\centering
\includegraphics[width=1.\textwidth]{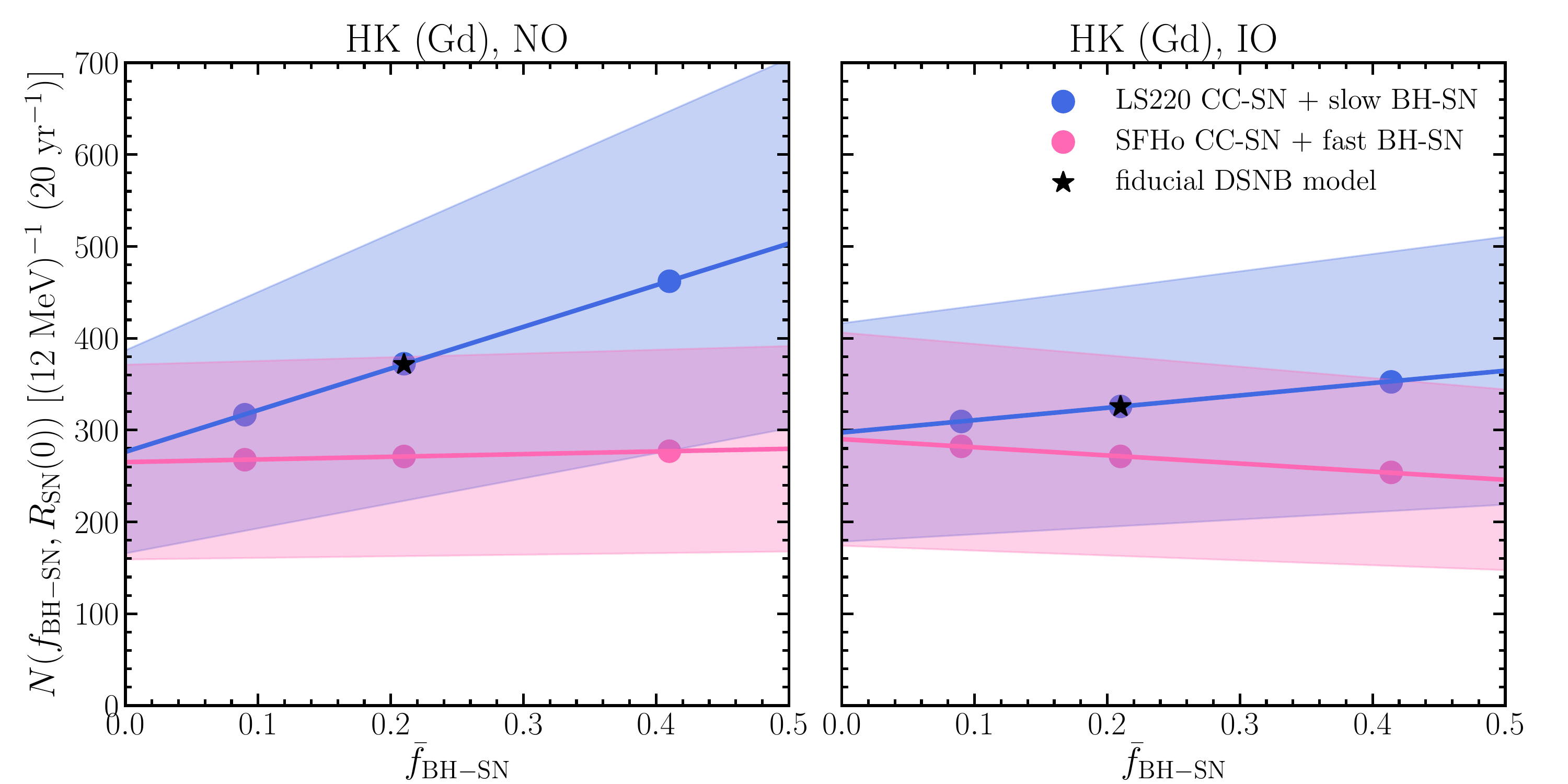}
 \caption{Number of expected events in the Hyper-Kamiokande (Gd)  detection region (12--24 MeV) as a function of $\bar{f}_{\mathrm{BH-SN}}$ for the LS220 EoS for CC-SN $+$slow BH-SN in blue and the SFHo EoS for CC-SN $+$ fast BH-SN in magenta in NO (on the left) and IO (on the right). The fiducial DSNB model is marked with a star. The bands refer to the uncertainties on the local supernova rate ($R_{\mathrm{SN}}(0) \in [1.25-0.5, 1.25+0.5] \times 10^{-4}$~Mpc$^{-3}$~yr$^{-1}$). The overlaps between the blue and the magenta band indicates a similar DSNB signal for different model parameters. }
 \label{fig:fBHtrend}
\end{figure}

As a consequence of the degeneracy among the DSNB model parameters, and its dependence on the mass ordering, the expected trend that the number of expected events increases as $f_{\mathrm{BH-SN}}$ increases (see Fig.~\ref{fig:DSNBdeg}) does not hold in IO for the SFHo CC-SN $+$fast BH-SN cases. This is clearly shown in Fig.~\ref{fig:fBHtrend} where we report the expected number of events in  Hyper-Kamiokande enriched with Gd in NO (on the left) and in IO (on the right) within the selected detection region (12--24 MeV). 
The event rate for LS220 CC-SN EoS $+$slow BH-SN is shown in blue and the one for SFHo CC-SN EoS $+$ fast BH-SN is plotted in magenta. The bands refer to the uncertainties on the local supernova rate. One can clearly see that the same event rate (and a similar DSNB signal as a function of the energy) can be obtained for different values of the model parameters. This indicates that without precise measurement of $\bar R_{SN}(0)$ from other channels than neutrinos, it will be impossible to determine $\bar f_{BH-SN}$.

\begin{figure}[t]
\centering
\includegraphics[width=1.\textwidth]{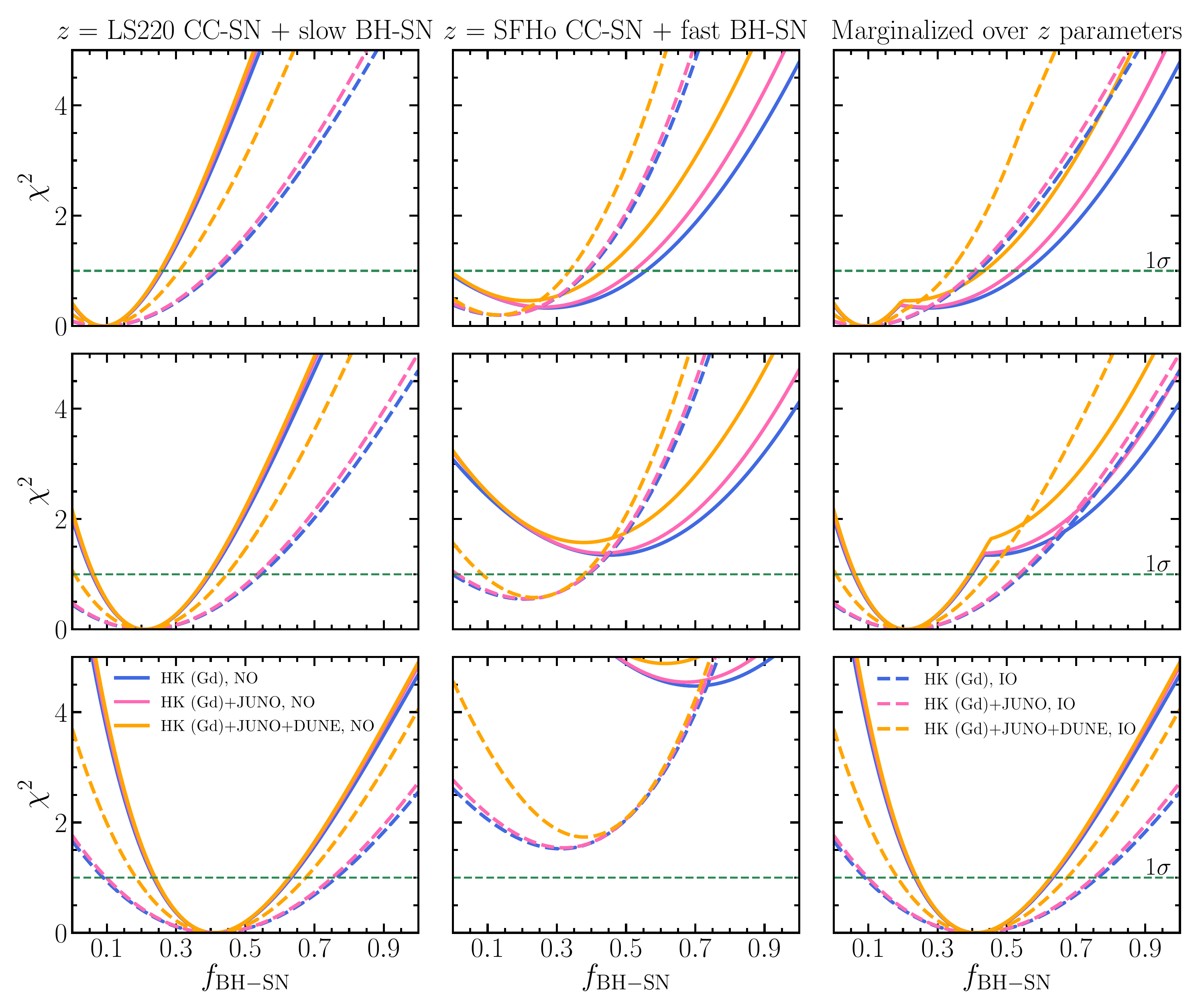}
 \caption{$\chi^2$ dependence on $f_{\mathrm{BH-SN}}$ for NO (solid lines) and IO (dashed lines) for two different $z$ (left panels: LS220 EoS for CC-SN $+$ slow SN-BH; middle panels: SFHo EoS for CC-SN $+$ fast SN-BH) and minimization with respect to the $z$ (right panels). From top to bottom: $\bar{f}_{\mathrm{BH-SN}}=  0.09, 0.21, 0.41$. In NO the second local minimum appearing in the combined $\chi^2$ is determined by the dependence of the $\chi^2$ on $z$  as a function of $f_{\mathrm{BH-SN}}$. The second local minimum does not appear in IO.
} 
 \label{fig:chisq_parabolas}
\end{figure}
In the light of the degeneracy among the model parameters and in order to better motivate the shape with two local minima in the $\chi^2$ in Fig.~\ref{fig:fBH_chisq}, Fig.~\ref{fig:chisq_parabolas} shows the $\chi^2$ dependence as a function of $f_{\mathrm{BH-SN}}$ in NO (solid lines) and IO (dashed lines) for the LS220 EoS for CC-SN $+$ slow SN-BH and the SFHo EoS for CC-SN $+$ fast SN-BH cases in  the left and middle columns respectively\footnote{Other permutations of the EoS for the CC-SN and SN-BH yield similar results and are not shown here}). The resultant $\chi^2$ with minimization with respect to the CC-SN EoS and BH-SN mass accretion rate ($z$) is shown in the right column. One can see that  in NO the second local  minimum appearing in the combined $\chi^2$ is determined by the dependence of the $\chi^2$ on the EoS CC-SN $+$ BH-SN accretion rate as a function of $f_{\mathrm{BH-SN}}$. However, the second minimum does not appear in IO.

\begin{figure}[t]
\centering
\includegraphics[width=1.\textwidth]{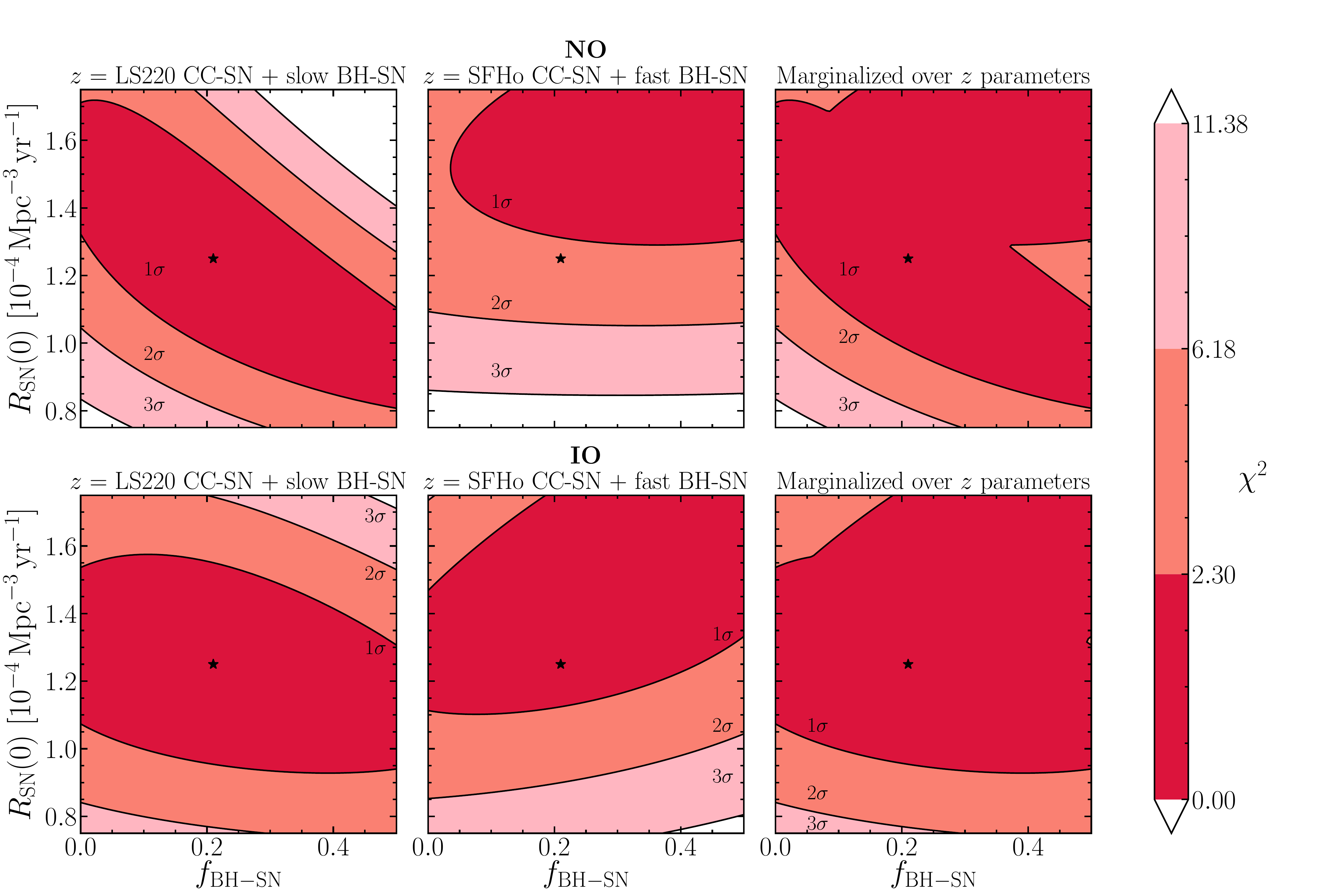}
 \caption{Contour plot of the allowed $\chi^2$ in the $(f_{\mathrm{BH-SN}}, R_{\mathrm{SN}})$ plane for NO (top panels) and IO (bottom panels) and for LS220 $+$slow EoS (left panels) and SFHo$+$fast EoS (middle panels), and minimized with respect to $z$ (right panels). The irregular contours appearing in the panels on the right are a direct consequence of the minimization with respect to $z$ and the degeneracies of the model parameters.}
  
  \label{fig:2Dchisqdeg}
\end{figure}

Figure~\ref{fig:2Dchisqdeg} shows the correspondent 2D $\chi^2$ analysis in the $(f_{\mathrm{BH-SN}}, R_{\mathrm{SN}})$ plane for NO (top panels) and IO (bottom panels) and for LS220 $+$slow EoS (left panels), SFHo$+$fast EoS (middle panels), and marginalized with respect to $z$ (right panels). 
The irregular shapes appearing in the $1 \sigma$ contours  for the $\chi^2$ statistics minimized with respect to $z$ (right panels) are due to the different dependencies of the allowed regions as a function of $z$ (see left and middle panels).

\begin{figure}[t]
\centering
\includegraphics[width=1.\textwidth]{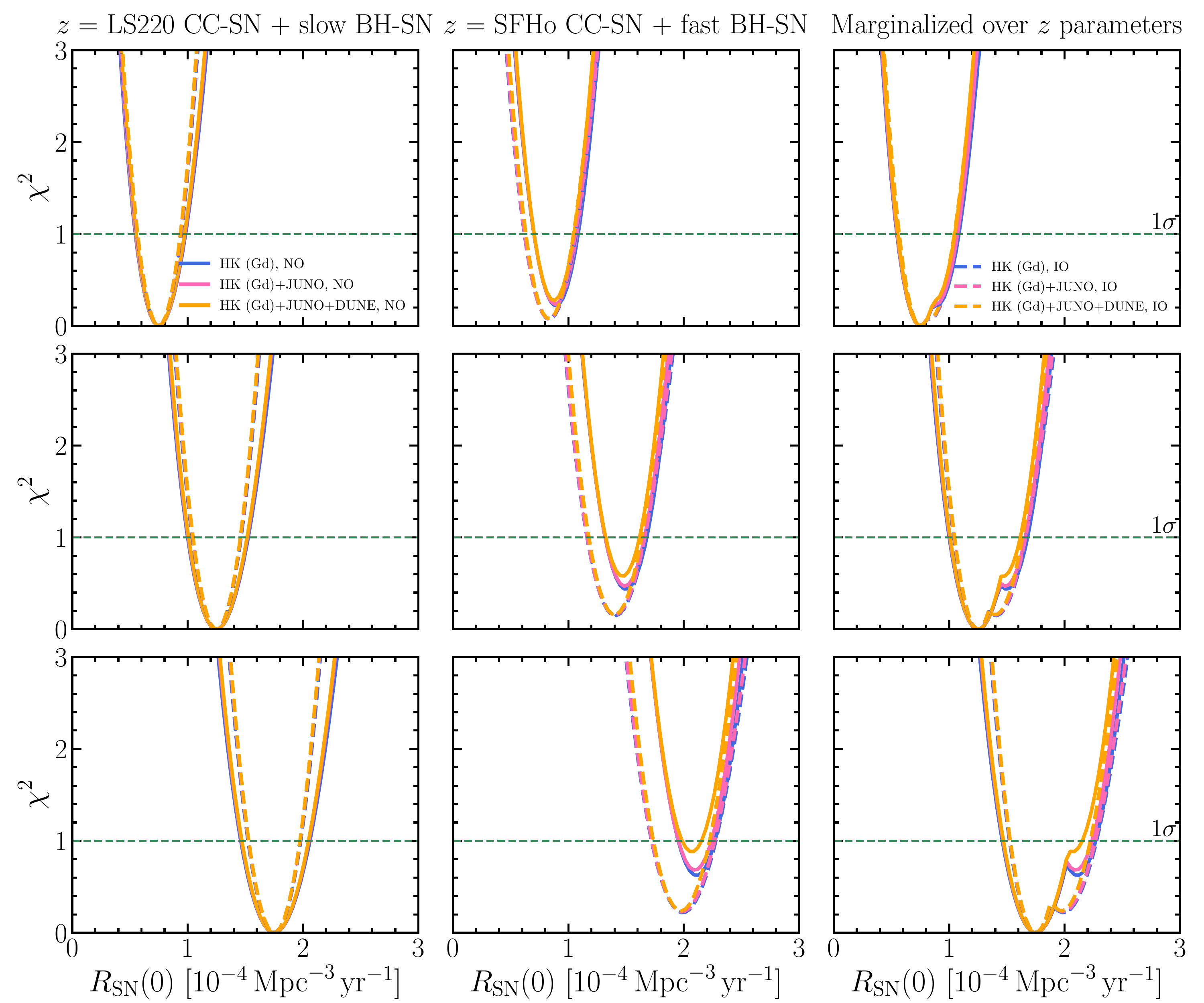}
 \caption{$\chi^2$ dependence on $\bar{R}_{\mathrm{SN}}$ for NO (solid lines) and IO (dashed lines) for two different $z$ (left panel: LS220 EoS for CC-SN $+$ slow SN-BH; middle panel: SFHo EoS for CC-SN $+$ fast SN-BH; right panel: $\chi^2$ with minimization with respect to the $z$). From top to bottom: $\bar{R}_{\mathrm{SN}}= 0.75, 1.25, 1.75$. The second local minimum in the $\chi^2$ negligibly affects the shape of the $\chi^2$ marginalized with respect to $z$.} 
\label{fig:chisq_noparabolas}
\end{figure}

Figure~\ref{fig:chisq_noparabolas} shows the $\chi^2$ dependence of the local SN rate as a function of $R_{\mathrm{SN}}(0)$ in NO (solid lines) and IO (dashed lines) for the LS220 EoS for CC-SN $+$ slow SN-BH and the SFHo CC-SN $+$ fast SN-BH cases in  the left and middle columns respectively and marginalized with respect to $z$ on the right. The second local minimum in the $\chi^2$ is not pronounced as in the case of  Fig.~\ref{fig:chisq_parabolas} and therefore it negligibly affects the shape of the $\chi^2$ in the plots on the right column. 

\begin{figure}[t]
\centering
\includegraphics[width=1.\textwidth]{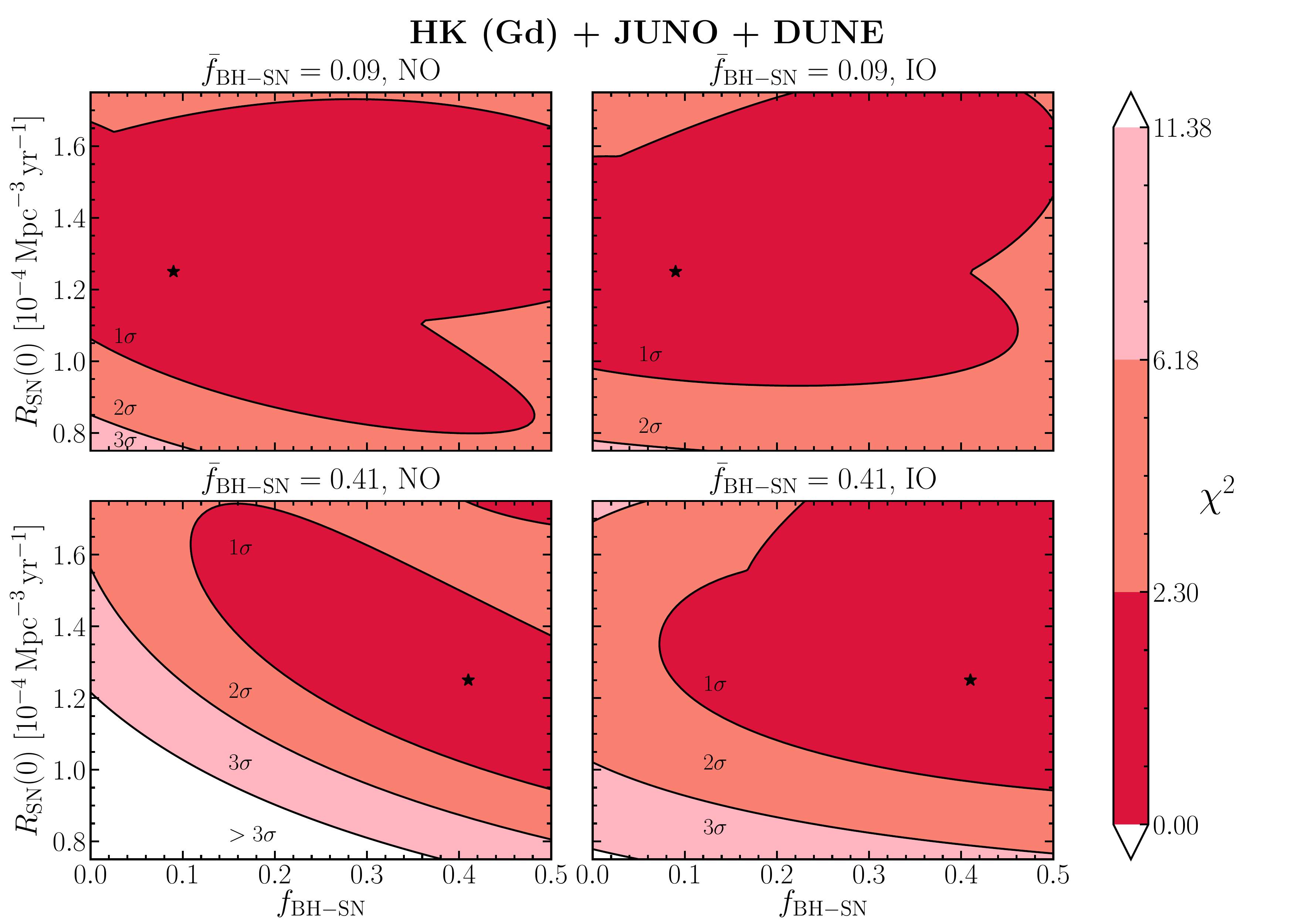}
 \caption{Contour plot of the local $R_{\mathrm{SN}}(0)$ as a function of $f_{\mathrm{BH-SN}}$. The fiducial DSNB model is marked by a star. The upper (bottom) panels refer to the case in which the NC atmospheric background is neglected in Hyper-Kamiokande (Gd) and $\bar{f}_{\mathrm{BH-SN}} = 9\%$ ($\bar{f}_{\mathrm{BH-SN}} = 41\%$). The left (right) panels have been obtained assuming the NO (IO). The three contours indicate the $1, 2$ and $3 \sigma$ confidence levels, respectively for 2 d.o.f. As $\bar{f}_{\mathrm{BH-SN}}$ increases the event rate is larger and the allowed parameter space shrinks.}  \label{fig:contours2}
\end{figure}

Figure~\ref{fig:contours2} shows the contour plot of the local $R_{\mathrm{SN}}(0)$ as a function of $f_{\mathrm{BH-SN}}$ for the DSNB fiducial model and $\bar{f}_{\mathrm{BH-SN}} = 9\%$ on the top and  $\bar{f}_{\mathrm{BH-SN}} = 41\%$ on the bottom. Comparing this figure with Fig.~\ref{fig:contours}, one can see that as $\bar{f}_{\mathrm{BH-SN}}$ increases the constrained parameter space shrinks as the DSNB even rate increases.

\bibliographystyle{JHEP}
\bibliography{DSNB}

\providecommand{\href}[2]{#2}\begingroup\raggedright\begin{thebibliography}{10}

\bibitem{Janka:2017vcp}
H.~T. Janka, \emph{{Neutrino-driven Explosions}}, {\emph{\relax{\textrm{Author
  version of chapter for ``Handbook of Supernovae,'' edited by A.~Alsabti and
  P.~Murdin, Springer}}} (2017) },
  [\href{https://arxiv.org/abs/1702.08825}{{\tt 1702.08825}}].

\bibitem{Mirizzi:2015eza}
A.~Mirizzi, I.~Tamborra, H.-T. Janka, N.~Saviano, K.~Scholberg, R.~Bollig
  et~al., \emph{{Supernova Neutrinos: Production, Oscillations and Detection}},
  \href{http://dx.doi.org/10.1393/ncr/i2016-10120-8}{\emph{Riv. Nuovo Cim.}
  {\bf 39} (2016) 1--112}, [\href{https://arxiv.org/abs/1508.00785}{{\tt
  1508.00785}}].

\bibitem{Beacom:2010kk}
J.~F. Beacom, \emph{{The Diffuse Supernova Neutrino Background}},
  \href{http://dx.doi.org/10.1146/annurev.nucl.010909.083331}{\emph{Ann. Rev.
  Nucl. Part. Sci.} {\bf 60} (2010) 439--462},
  [\href{https://arxiv.org/abs/1004.3311}{{\tt 1004.3311}}].

\bibitem{Lunardini:2010ab}
C.~Lunardini, \emph{{Diffuse supernova neutrinos at underground laboratories}},
  \href{http://dx.doi.org/10.1016/j.astropartphys.2016.02.005}{\emph{Astropart.
  Phys.} {\bf 79} (2016) 49--77}, [\href{https://arxiv.org/abs/1007.3252}{{\tt
  1007.3252}}].

\bibitem{Lunardini:2009ya}
C.~Lunardini, \emph{{Diffuse neutrino flux from failed supernovae}},
  \href{http://dx.doi.org/10.1103/PhysRevLett.102.231101}{\emph{Phys. Rev.
  Lett.} {\bf 102} (2009) 231101}, [\href{https://arxiv.org/abs/0901.0568}{{\tt
  0901.0568}}].

\bibitem{Horiuchi:2017qja}
S.~{Horiuchi}, K.~{Sumiyoshi}, K.~{Nakamura}, T.~{Fischer}, A.~{Summa},
  T.~{Takiwaki} et~al., \emph{{Diffuse supernova neutrino background from
  extensive core-collapse simulations of 8-100 M$_{?}$ progenitors}},
  \href{http://dx.doi.org/10.1093/mnras/stx3271}{\emph{Mon. Not. Roy. Astron.
  Soc.} {\bf 475} (Mar., 2018) 1363--1374},
  [\href{https://arxiv.org/abs/1709.06567}{{\tt 1709.06567}}].

\bibitem{Nakazato:2015rya}
K.~Nakazato, E.~Mochida, Y.~Niino and H.~Suzuki, \emph{{Spectrum of the
  Supernova Relic Neutrino Background and Metallicity Evolution of Galaxies}},
  \href{http://dx.doi.org/10.1088/0004-637X/804/1/75}{\emph{Astrophys. J.} {\bf
  804} (2015) 75}, [\href{https://arxiv.org/abs/1503.01236}{{\tt 1503.01236}}].

\bibitem{Nakazato:2013maa}
K.~Nakazato, \emph{{Imprint of Explosion Mechanism on Supernova Relic
  Neutrinos}}, \href{http://dx.doi.org/10.1103/PhysRevD.88.083012}{\emph{Phys.
  Rev.} {\bf D88} (2013) 083012}, [\href{https://arxiv.org/abs/1306.4526}{{\tt
  1306.4526}}].

\bibitem{Lunardini:2012ne}
C.~Lunardini and I.~Tamborra, \emph{{Diffuse supernova neutrinos: oscillation
  effects, stellar cooling and progenitor mass dependence}},
  \href{http://dx.doi.org/10.1088/1475-7516/2012/07/012}{\emph{JCAP} {\bf 1207}
  (2012) 012}, [\href{https://arxiv.org/abs/1205.6292}{{\tt 1205.6292}}].

\bibitem{Ando:2004hc}
S.~Ando and K.~Sato, \emph{{Relic neutrino background from cosmological
  supernovae}}, \href{http://dx.doi.org/10.1088/1367-2630/6/1/170}{\emph{New J.
  Phys.} {\bf 6} (2004) 170},
  [\href{https://arxiv.org/abs/astro-ph/0410061}{{\tt astro-ph/0410061}}].

\bibitem{Chakraborty:2008zp}
S.~Chakraborty, S.~Choubey, B.~Dasgupta and K.~Kar, \emph{{Effect of Collective
  Flavor Oscillations on the Diffuse Supernova Neutrino Background}},
  \href{http://dx.doi.org/10.1088/1475-7516/2008/09/013}{\emph{JCAP} {\bf 0809}
  (2008) 013}, [\href{https://arxiv.org/abs/0805.3131}{{\tt 0805.3131}}].

\bibitem{Chakraboty:2010sz}
S.~Chakraborty, S.~Choubey and K.~Kar, \emph{{On the Observability of
  Collective Flavor Oscillations in Diffuse Supernova Neutrino Background}},
  \href{http://dx.doi.org/10.1016/j.physletb.2011.06.089}{\emph{Phys. Lett.}
  {\bf B702} (2011) 209--215}, [\href{https://arxiv.org/abs/1006.3756}{{\tt
  1006.3756}}].

\bibitem{Ertl:2015rga}
T.~Ertl, H.~T. Janka, S.~E. Woosley, T.~Sukhbold and M.~Ugliano, \emph{{A
  two-parameter criterion for classifying the explodability of massive stars by
  the neutrino-driven mechanism}},
  \href{http://dx.doi.org/10.3847/0004-637X/818/2/124}{\emph{Astrophys. J.}
  {\bf 818} (2016) 124}, [\href{https://arxiv.org/abs/1503.07522}{{\tt
  1503.07522}}].

\bibitem{Sukhbold:2015wba}
T.~Sukhbold, T.~Ertl, S.~E. Woosley, J.~M. Brown and H.~T. Janka,
  \emph{{Core-Collapse Supernovae from 9 to 120 Solar Masses Based on
  Neutrino-powered Explosions}},
  \href{http://dx.doi.org/10.3847/0004-637X/821/1/38}{\emph{Astrophys. J.} {\bf
  821} (2016) 38}, [\href{https://arxiv.org/abs/1510.04643}{{\tt 1510.04643}}].

\bibitem{Adams:2016hit}
S.~M. Adams, C.~S. Kochanek, J.~R. Gerke and K.~Z. Stanek, \emph{{The search
  for failed supernovae with the Large Binocular Telescope: constraints from 7
  yr of data}}, \href{http://dx.doi.org/10.1093/mnras/stx898}{\emph{Mon. Not.
  Roy. Astron. Soc.} {\bf 469} (2017) 1445--1455},
  [\href{https://arxiv.org/abs/1610.02402}{{\tt 1610.02402}}].

\bibitem{Adams:2016ffj}
S.~M. Adams, C.~S. Kochanek, J.~R. Gerke, K.~Z. Stanek and X.~Dai, \emph{{The
  search for failed supernovae with the Large Binocular Telescope: confirmation
  of a disappearing star}},
  \href{http://dx.doi.org/10.1093/mnras/stx816}{\emph{Mon. Not. Roy. Astron.
  Soc.} {\bf 468} (2017) 4968--4981},
  [\href{https://arxiv.org/abs/1609.01283}{{\tt 1609.01283}}].

\bibitem{Gerke:2014ooa}
J.~R. Gerke, C.~S. Kochanek and K.~Z. Stanek, \emph{{The Search for Failed
  Supernovae with The Large Binocular Telescope: First Candidates}},
  \href{http://dx.doi.org/10.1093/mnras/stv776}{\emph{Mon. Not. Roy. Astron.
  Soc.} {\bf 450} (2015) 3289--3305},
  [\href{https://arxiv.org/abs/1411.1761}{{\tt 1411.1761}}].

\bibitem{Kochanek:2008mp}
C.~S. Kochanek et~al., \emph{{A Survey About Nothing: Monitoring a Million
  Supergiants for Failed Supernovae}},
  \href{http://dx.doi.org/10.1086/590053}{\emph{Astrophys. J.} {\bf 684} (2008)
  1336--1342}, [\href{https://arxiv.org/abs/0802.0456}{{\tt 0802.0456}}].

\bibitem{Horiuchi:2011zz}
S.~Horiuchi, J.~F. Beacom, C.~S. Kochanek, J.~L. Prieto, K.~Z. Stanek and T.~A.
  Thompson, \emph{{The Cosmic Core-collapse Supernova Rate does not match the
  Massive-Star Formation Rate}},
  \href{http://dx.doi.org/10.1088/0004-637X/738/2/154}{\emph{Astrophys. J.}
  {\bf 738} (2011) 154--169}, [\href{https://arxiv.org/abs/1102.1977}{{\tt
  1102.1977}}].

\bibitem{Bays:2011si}
{\scshape Super-Kamiokande} collaboration, K.~Bays et~al., \emph{{Supernova
  Relic Neutrino Search at Super-Kamiokande}},
  \href{http://dx.doi.org/10.1103/PhysRevD.85.052007}{\emph{Phys. Rev.} {\bf
  D85} (2012) 052007}, [\href{https://arxiv.org/abs/1111.5031}{{\tt
  1111.5031}}].

\bibitem{Zhang:2013tua}
{\scshape Super-Kamiokande} collaboration, H.~Zhang et~al., \emph{{Supernova
  Relic Neutrino Search with Neutron Tagging at Super-Kamiokande-IV}},
  \href{http://dx.doi.org/10.1016/j.astropartphys.2014.05.004}{\emph{Astropart.
  Phys.} {\bf 60} (2015) 41--46}, [\href{https://arxiv.org/abs/1311.3738}{{\tt
  1311.3738}}].

\bibitem{Zhang:2015zla}
{\scshape Super-Kamiokande} collaboration, Y.~Zhang, \emph{{Search for
  Supernova Relic Neutrinos with 2.2 MeV Gamma Tagging at
  Super-Kamiokande-IV}},
  \href{http://dx.doi.org/10.1016/j.phpro.2014.12.079}{\emph{Phys. Procedia}
  {\bf 61} (2015) 384--391}.

\bibitem{Beacom:2003nk}
J.~F. Beacom and M.~R. Vagins, \emph{{GADZOOKS! Anti-neutrino spectroscopy with
  large water Cherenkov detectors}},
  \href{http://dx.doi.org/10.1103/PhysRevLett.93.171101}{\emph{Phys. Rev.
  Lett.} {\bf 93} (2004) 171101},
  [\href{https://arxiv.org/abs/hep-ph/0309300}{{\tt hep-ph/0309300}}].

\bibitem{Horiuchi:2008jz}
S.~Horiuchi, J.~F. Beacom and E.~Dwek, \emph{{The Diffuse Supernova Neutrino
  Background is detectable in Super-Kamiokande}},
  \href{http://dx.doi.org/10.1103/PhysRevD.79.083013}{\emph{Phys. Rev.} {\bf
  D79} (2009) 083013}, [\href{https://arxiv.org/abs/0812.3157}{{\tt
  0812.3157}}].

\bibitem{Watanabe:2008ru}
{\scshape Super-Kamiokande} collaboration, H.~Watanabe et~al., \emph{{First
  Study of Neutron Tagging with a Water Cherenkov Detector}},
  \href{http://dx.doi.org/10.1016/j.astropartphys.2009.03.002}{\emph{Astropart.
  Phys.} {\bf 31} (2009) 320--328},
  [\href{https://arxiv.org/abs/0811.0735}{{\tt 0811.0735}}].

\bibitem{Xu:2016cfv}
{\scshape Super-Kamiokande} collaboration, C.~Xu, \emph{{Current status of
  SK-Gd project and EGADS}},
  \href{http://dx.doi.org/10.1088/1742-6596/718/6/062070}{\emph{J. Phys. Conf.
  Ser.} {\bf 718} (2016) 062070}.

\bibitem{Sekiya:2016xji}
{\scshape SUPER-KAMIOKANDE} collaboration, H.~Sekiya, \emph{{The
  Super-Kamiokande Gadolinium Project}}, {\emph{PoS} {\bf ICHEP2016} (2016)
  982}.

\bibitem{Hyper-Kamiokande:2016dsw}
{\scshape Hyper-Kamiokande} collaboration, K.~Abe et~al.,
  \emph{{Hyper-Kamiokande Design Report}}, .

\bibitem{An:2015jdp}
{\scshape JUNO} collaboration, F.~An et~al., \emph{{Neutrino Physics with
  JUNO}}, \href{http://dx.doi.org/10.1088/0954-3899/43/3/030401}{\emph{J.
  Phys.} {\bf G43} (2016) 030401},
  [\href{https://arxiv.org/abs/1507.05613}{{\tt 1507.05613}}].

\bibitem{Wei:2016vjd}
H.~Wei, Z.~Wang and S.~Chen, \emph{{Discovery potential for supernova relic
  neutrinos with slow liquid scintillator detectors}},
  \href{http://dx.doi.org/10.1016/j.physletb.2017.03.071}{\emph{Phys. Lett.}
  {\bf B769} (2017) 255--261}, [\href{https://arxiv.org/abs/1607.01671}{{\tt
  1607.01671}}].

\bibitem{Priya:2017bmm}
A.~Priya and C.~Lunardini, \emph{{Diffuse neutrinos from luminous and dark
  supernovae: prospects for upcoming detectors at the $O$(10) kt scale}},
  \href{http://dx.doi.org/10.1088/1475-7516/2017/11/031}{\emph{JCAP} {\bf 1711}
  (2017) 031}, [\href{https://arxiv.org/abs/1705.02122}{{\tt 1705.02122}}].

\bibitem{Acciarri:2016crz}
{\scshape DUNE} collaboration, R.~Acciarri et~al., \emph{{Long-Baseline
  Neutrino Facility (LBNF) and Deep Underground Neutrino Experiment (DUNE)}},
  \href{https://arxiv.org/abs/1601.05471}{{\tt 1601.05471}}.

\bibitem{Garc:SN}
``Garching core-collapse supernova archive.''
  {\url{https://wwwmpa.mpa-garching.mpg.de/ccsnarchive/}}.

\bibitem{Lattimer:1991nc}
J.~M. Lattimer and F.~D. Swesty, \emph{{A Generalized equation of state for
  hot, dense matter}},
  \href{http://dx.doi.org/10.1016/0375-9474(91)90452-C}{\emph{Nucl. Phys.} {\bf
  A535} (1991) 331--376}.

\bibitem{Steiner:2012rk}
A.~W. Steiner, M.~Hempel and T.~Fischer, \emph{{Core-collapse supernova
  equations of state based on neutron star observations}},
  \href{http://dx.doi.org/10.1088/0004-637X/774/1/17}{\emph{Astrophys. J.} {\bf
  774} (2013) 17}, [\href{https://arxiv.org/abs/1207.2184}{{\tt 1207.2184}}].

\bibitem{Keil:2002in}
M.~T. Keil, G.~G. Raffelt and H.-T. Janka, \emph{{Monte Carlo study of
  supernova neutrino spectra formation}},
  \href{http://dx.doi.org/10.1086/375130}{\emph{Astrophys. J.} {\bf 590} (2003)
  971--991}, [\href{https://arxiv.org/abs/astro-ph/0208035}{{\tt
  astro-ph/0208035}}].

\bibitem{Tamborra:2012ac}
I.~Tamborra, B.~M{\"u}ller, L.~H{\"u}depohl, H.-T. Janka and G.~Raffelt,
  \emph{{High-resolution supernova neutrino spectra represented by a simple
  fit}}, \href{http://dx.doi.org/10.1103/PhysRevD.86.125031}{\emph{Phys. Rev.}
  {\bf D86} (2012) 125031}, [\href{https://arxiv.org/abs/1211.3920}{{\tt
  1211.3920}}].

\bibitem{Chakraborty:2016yeg}
S.~Chakraborty, R.~Hansen, I.~Izaguirre and G.~Raffelt, \emph{{Collective
  neutrino flavor conversion: Recent developments}},
  \href{http://dx.doi.org/10.1016/j.nuclphysb.2016.02.012}{\emph{Nucl. Phys.}
  {\bf B908} (2016) 366--381}, [\href{https://arxiv.org/abs/1602.02766}{{\tt
  1602.02766}}].

\bibitem{Duan:2010bg}
H.~Duan, G.~M. Fuller and Y.-Z. Qian, \emph{{Collective Neutrino
  Oscillations}},
  \href{http://dx.doi.org/10.1146/annurev.nucl.012809.104524}{\emph{Ann. Rev.
  Nucl. Part. Sci.} {\bf 60} (2010) 569--594},
  [\href{https://arxiv.org/abs/1001.2799}{{\tt 1001.2799}}].

\bibitem{EstebanPretel:2008ni}
A.~Esteban-Pretel, A.~Mirizzi, S.~Pastor, R.~Tom\`as, G.~G. Raffelt, P.~D.
  Serpico et~al., \emph{{Role of dense matter in collective supernova neutrino
  transformations}},
  \href{http://dx.doi.org/10.1103/PhysRevD.78.085012}{\emph{Phys. Rev.} {\bf
  D78} (2008) 085012}, [\href{https://arxiv.org/abs/0807.0659}{{\tt
  0807.0659}}].

\bibitem{Tamborra:2017ubu}
I.~Tamborra, L.~H{\"u}depohl, G.~Raffelt and H.-T. Janka,
  \emph{{Flavor-dependent neutrino angular distribution in core-collapse
  supernovae}},
  \href{http://dx.doi.org/10.3847/1538-4357/aa6a18}{\emph{Astrophys. J.} {\bf
  839} (2017) 132}, [\href{https://arxiv.org/abs/1702.00060}{{\tt
  1702.00060}}].

\bibitem{Izaguirre:2016gsx}
I.~Izaguirre, G.~Raffelt and I.~Tamborra, \emph{{Fast Pairwise Conversion of
  Supernova Neutrinos: A Dispersion-Relation Approach}},
  \href{http://dx.doi.org/10.1103/PhysRevLett.118.021101}{\emph{Phys. Rev.
  Lett.} {\bf 118} (2017) 021101},
  [\href{https://arxiv.org/abs/1610.01612}{{\tt 1610.01612}}].

\bibitem{Sawyer:2015dsa}
R.~F. Sawyer, \emph{{Neutrino cloud instabilities just above the neutrino
  sphere of a supernova}},
  \href{http://dx.doi.org/10.1103/PhysRevLett.116.081101}{\emph{Phys. Rev.
  Lett.} {\bf 116} (2016) 081101},
  [\href{https://arxiv.org/abs/1509.03323}{{\tt 1509.03323}}].

\bibitem{Sawyer:2005jk}
R.~F. Sawyer, \emph{{Speed-up of neutrino transformations in a supernova
  environment}},
  \href{http://dx.doi.org/10.1103/PhysRevD.72.045003}{\emph{Phys. Rev.} {\bf
  D72} (2005) 045003}, [\href{https://arxiv.org/abs/hep-ph/0503013}{{\tt
  hep-ph/0503013}}].

\bibitem{Mikheev:1986if}
S.~P. Mikheev and A.~{\relax Yu}. Smirnov, \emph{{Neutrino Oscillations in a
  Variable Density Medium and Neutrino Bursts Due to the Gravitational Collapse
  of Stars}}, {\emph{Sov. Phys. JETP} {\bf 64} (1986) 4--7},
  [\href{https://arxiv.org/abs/0706.0454}{{\tt 0706.0454}}].

\bibitem{1978PhRvD..17.2369W}
L.~{Wolfenstein}, \emph{{Neutrino oscillations in matter}},
  \href{http://dx.doi.org/10.1103/PhysRevD.17.2369}{\emph{Phys. Rev. D} {\bf
  17} (May, 1978) 2369--2374}.

\bibitem{1985YaFiz..42.1441M}
S.~P. {Mikheyev} and A.~{\relax Yu}. {Smirnov}, \emph{{Resonance enhancement of
  oscillations in matter and solar neutrino spectroscopy}}, {\emph{Yadernaya
  Fizika} {\bf 42} (1985) 1441--1448}.

\bibitem{Dighe:1999bi}
A.~S. Dighe and A.~{\relax Yu}. Smirnov, \emph{{Identifying the neutrino mass
  spectrum from the neutrino burst from a supernova}},
  \href{http://dx.doi.org/10.1103/PhysRevD.62.033007}{\emph{Phys. Rev.} {\bf
  D62} (2000) 033007}, [\href{https://arxiv.org/abs/hep-ph/9907423}{{\tt
  hep-ph/9907423}}].

\bibitem{Capozzi:2017ipn}
F.~Capozzi, E.~Di~Valentino, E.~Lisi, A.~Marrone, A.~Melchiorri and A.~Palazzo,
  \emph{{Global constraints on absolute neutrino masses and their ordering}},
  \href{http://dx.doi.org/10.1103/PhysRevD.95.096014}{\emph{Phys. Rev.} {\bf
  D95} (2017) 096014}, [\href{https://arxiv.org/abs/1703.04471}{{\tt
  1703.04471}}].

\bibitem{Patterson:2015xja}
R.~B. Patterson, \emph{{Prospects for Measurement of the Neutrino Mass
  Hierarchy}},
  \href{http://dx.doi.org/10.1146/annurev-nucl-102014-021916}{\emph{Ann. Rev.
  Nucl. Part. Sci.} {\bf 65} (2015) 177--192},
  [\href{https://arxiv.org/abs/1506.07917}{{\tt 1506.07917}}].

\bibitem{1955ApJ...121..161S}
E.~E. {Salpeter}, \emph{{The Luminosity Function and Stellar Evolution.}},
  \href{http://dx.doi.org/10.1086/145971}{\emph{Astrophys. J.} {\bf 121} (Jan.,
  1955) 161}.

\bibitem{Yuksel:2008cu}
H.~Yuksel, M.~D. Kistler, J.~F. Beacom and A.~M. Hopkins, \emph{{Revealing the
  High-Redshift Star Formation Rate with Gamma-Ray Bursts}},
  \href{http://dx.doi.org/10.1086/591449}{\emph{Astrophys. J.} {\bf 683} (2008)
  L5--L8}, [\href{https://arxiv.org/abs/0804.4008}{{\tt 0804.4008}}].

\bibitem{Strolger:2015kra}
L.-G. Strolger et~al., \emph{{The Rate of Core Collapse Supernovae to Redshift
  2.5 From The CANDELS and CLASH Supernova Surveys}},
  \href{http://dx.doi.org/10.1088/0004-637X/813/2/93}{\emph{Astrophys. J.} {\bf
  813} (2015) 93}, [\href{https://arxiv.org/abs/1509.06574}{{\tt 1509.06574}}].

\bibitem{Madau:2014bja}
P.~Madau and M.~Dickinson, \emph{{Cosmic Star Formation History}},
  \href{http://dx.doi.org/10.1146/annurev-astro-081811-125615}{\emph{Ann. Rev.
  Astron. Astrophys.} {\bf 52} (2014) 415--486},
  [\href{https://arxiv.org/abs/1403.0007}{{\tt 1403.0007}}].

\bibitem{Lien:2010yb}
A.~Lien, B.~D. Fields and J.~F. Beacom, \emph{{Synoptic Sky Surveys and the
  Diffuse Supernova Neutrino Background: Removing Astrophysical Uncertainties
  and Revealing Invisible Supernovae}},
  \href{http://dx.doi.org/10.1103/PhysRevD.81.083001}{\emph{Phys. Rev.} {\bf
  D81} (2010) 083001}, [\href{https://arxiv.org/abs/1001.3678}{{\tt
  1001.3678}}].

\bibitem{Mattila:2012zr}
S.~Mattila et~al., \emph{{Core-collapse supernovae missed by optical surveys}},
  \href{http://dx.doi.org/10.1088/0004-637X/756/2/111}{\emph{Astrophys. J.}
  {\bf 756} (2012) 111}, [\href{https://arxiv.org/abs/1206.1314}{{\tt
  1206.1314}}].

\bibitem{Taylor:2014rlo}
M.~Taylor et~al., \emph{{The Core Collapse Supernova Rate from the SDSS-II
  Supernova Survey}},
  \href{http://dx.doi.org/10.1088/0004-637X/792/2/135}{\emph{Astrophys. J.}
  {\bf 792} (2014) 135}, [\href{https://arxiv.org/abs/1407.0999}{{\tt
  1407.0999}}].

\bibitem{Botticella:2011nd}
M.~T. Botticella, S.~J. Smartt, R.~C. Kennicutt, Jr., E.~Cappellaro, M.~Sereno
  and J.~C. Lee, \emph{{A comparison between star formation rate diagnostics
  and rate of core collapse supernovae within 11 Mpc}},
  \href{http://dx.doi.org/10.1051/0004-6361/201117343}{\emph{Astron.
  Astrophys.} {\bf 537} (2012) A132},
  [\href{https://arxiv.org/abs/1111.1692}{{\tt 1111.1692}}].

\bibitem{Li:2010kc}
W.~Li et~al., \emph{{Nearby Supernova Rates from the Lick Observatory Supernova
  Search. II. The Observed Luminosity Functions and Fractions of Supernovae in
  a Complete Sample}},
  \href{http://dx.doi.org/10.1111/j.1365-2966.2011.18160.x}{\emph{Mon. Not.
  Roy. Astron. Soc.} {\bf 412} (2011) 1441},
  [\href{https://arxiv.org/abs/1006.4612}{{\tt 1006.4612}}].

\bibitem{Li:2010kd}
W.~Li et~al., \emph{{Nearby Supernova Rates from the Lick Observatory Supernova
  Search. III. The Rate-Size Relation, and the Rates as a Function of Galaxy
  Hubble Type and Colour}},
  \href{http://dx.doi.org/10.1111/j.1365-2966.2011.18162.x}{\emph{Mon. Not.
  Roy. Astron. Soc.} {\bf 412} (2011) 1473},
  [\href{https://arxiv.org/abs/1006.4613}{{\tt 1006.4613}}].

\bibitem{Cappellaro:2015qia}
E.~Cappellaro et~al., \emph{{Supernova rates from the SUDARE VST-OmegaCAM
  search - I. Rates per unit volume}},
  \href{http://dx.doi.org/10.1051/0004-6361/201526712}{\emph{Astron.
  Astrophys.} {\bf 584} (2015) A62},
  [\href{https://arxiv.org/abs/1509.04496}{{\tt 1509.04496}}].

\bibitem{Woosley:2002zz}
S.~E. Woosley, A.~Heger and T.~A. Weaver, \emph{{The evolution and explosion of
  massive stars}},
  \href{http://dx.doi.org/10.1103/RevModPhys.74.1015}{\emph{Rev. Mod. Phys.}
  {\bf 74} (2002) 1015--1071}.

\bibitem{Pejcha:2014wda}
O.~Pejcha and T.~A. Thompson, \emph{{The Landscape of the Neutrino Mechanism of
  Core-Collapse Supernovae: Neutron Star and Black Hole Mass Functions,
  Explosion Energies and Nickel Yields}},
  \href{http://dx.doi.org/10.1088/0004-637X/801/2/90}{\emph{Astrophys. J.} {\bf
  801} (2015) 90}, [\href{https://arxiv.org/abs/1409.0540}{{\tt 1409.0540}}].

\bibitem{Raithel:2017nlc}
{Raithel, Carolyn A. and Sukhbold, Tuguldur and \"Ozel, Feryal},
  \emph{{Confronting Models of Massive Star Evolution and Explosions with
  Remnant Mass Measurements}},
  \href{http://dx.doi.org/10.3847/1538-4357/aab09b}{\emph{Astrophys. J.} {\bf
  856} (2018) 35}, [\href{https://arxiv.org/abs/1712.00021}{{\tt 1712.00021}}].

\bibitem{Hidaka:2016zei}
J.~Hidaka, T.~Kajino and G.~J. Mathews, \emph{{Red-supergiant and Supernova
  Rate Problems: Implication for the Relic Supernova Neutrino Spectrum}},
  \href{http://dx.doi.org/10.3847/0004-637X/827/1/85}{\emph{Astrophys. J.} {\bf
  827} (2016) 85}.

\bibitem{Rao:2005ab}
S.~M. Rao, D.~A. Turnshek and D.~Nestor, \emph{{Damped lyman alpha systems at
  z<1.65: the expanded sdss hst sample}},
  \href{http://dx.doi.org/10.1086/498132}{\emph{Astrophys. J.} {\bf 636} (2006)
  610--630}, [\href{https://arxiv.org/abs/astro-ph/0509469}{{\tt
  astro-ph/0509469}}].

\bibitem{Strumia:2003zx}
A.~Strumia and F.~Vissani, \emph{{Precise quasielastic neutrino/nucleon
  cross-section}},
  \href{http://dx.doi.org/10.1016/S0370-2693(03)00616-6}{\emph{Phys. Lett.}
  {\bf B564} (2003) 42--54},
  [\href{https://arxiv.org/abs/astro-ph/0302055}{{\tt astro-ph/0302055}}].

\bibitem{Kunxian:2015ymr}
K.~Huang, \emph{{Measurement of the neutrino-oxygen neutral current
  quasi-elastic interaction cross-section by observing nuclear de-excitation
  $\gamma$-rays in the T2K experiment}}.
\newblock PhD thesis, Kyoto U., 2015.

\bibitem{Beacom:privcomm}
J.~F. Beacom. private communication.

\bibitem{Acciarri:2016ooe}
{\scshape DUNE} collaboration, R.~Acciarri et~al., \emph{{Long-Baseline
  Neutrino Facility (LBNF) and Deep Underground Neutrino Experiment (DUNE)}},
  \href{https://arxiv.org/abs/1601.02984}{{\tt 1601.02984}}.

\bibitem{Acciarri:2015uup}
{\scshape DUNE} collaboration, R.~Acciarri et~al., \emph{{Long-Baseline
  Neutrino Facility (LBNF) and Deep Underground Neutrino Experiment (DUNE)}},
  \href{https://arxiv.org/abs/1512.06148}{{\tt 1512.06148}}.

\bibitem{Martinez}
G.~Mart\'inez-Pinedo. private communication.

\bibitem{Ankowski:2016lab}
A.~Ankowski et~al., \emph{{Supernova Physics at DUNE}},  in \emph{{Supernova
  Physics at DUNE Blacksburg, Virginia, USA, March 11-12, 2016}}, 2016.
\newblock \href{https://arxiv.org/abs/1608.07853}{{\tt 1608.07853}}.

\bibitem{DUNEp}
I.~G. Botella, ``\emph{DUNE detector design with the emphasis on low-energy
  reconstruction}.'' {Supernova Physics at DUNE Workshop,
  \url{http://indico.phys.vt.edu/event/32/session/6/contribution/6}}, 2016.

\bibitem{Cocco:2004ac}
A.~G. Cocco, A.~Ereditato, G.~Fiorillo, G.~Mangano and V.~Pettorino,
  \emph{{Supernova relic neutrinos in liquid argon detectors}},
  \href{http://dx.doi.org/10.1088/1475-7516/2004/12/002}{\emph{JCAP} {\bf 0412}
  (2004) 002}, [\href{https://arxiv.org/abs/hep-ph/0408031}{{\tt
  hep-ph/0408031}}].

\bibitem{Fogli:2002pt}
G.~L. Fogli, E.~Lisi, A.~Marrone, D.~Montanino and A.~Palazzo, \emph{{Getting
  the most from the statistical analysis of solar neutrino oscillations}},
  \href{http://dx.doi.org/10.1103/PhysRevD.66.053010}{\emph{Phys. Rev.} {\bf
  D66} (2002) 053010}, [\href{https://arxiv.org/abs/hep-ph/0206162}{{\tt
  hep-ph/0206162}}].

\bibitem{Acciarri:2014isz}
{\scshape ArgoNeuT} collaboration, R.~Acciarri et~al., \emph{{Measurements of
  Inclusive Muon Neutrino and Antineutrino Charged Current Differential Cross
  Sections on Argon in the NuMI Antineutrino Beam}},
  \href{http://dx.doi.org/10.1103/PhysRevD.89.112003}{\emph{Phys. Rev.} {\bf
  D89} (2014) 112003}, [\href{https://arxiv.org/abs/1404.4809}{{\tt
  1404.4809}}].

\bibitem{LArp}
N.~Dokania, ``\emph{CAPTAIN: Current Neutron and Future Stopped Pion Neutrino
  Measurements}.'' {Meeting of the APS Division of Particles and Fields,
  \url{https://indico.fnal.gov/event/11999/contribution/362}}, 2017.

\bibitem{Burgio:2018mcr}
G.~F. Burgio and A.~F. Fantina, \emph{{Nuclear Equation of state for Compact
  Stars and Supernovae}},  \href{https://arxiv.org/abs/1804.03020}{{\tt
  1804.03020}}.

\bibitem{Capozzi:2018ubv}
F.~Capozzi, E.~Lisi, A.~Marrone and A.~Palazzo, \emph{{Current unknowns in the
  three neutrino framework}},  \href{https://arxiv.org/abs/1804.09678}{{\tt
  1804.09678}}.

\end{thebibliography}\endgroup

\end{document}